\documentclass[12pt, letterpaper]{JHEP3}
\pdfoutput=1

\usepackage{amsmath}
\usepackage{amssymb}
\usepackage{amsfonts}
\usepackage{lscape, graphicx}
\usepackage{cite}       
\usepackage{bm}         
\usepackage{verbatim}

\usepackage[all]{xy}
\advance\parskip 1.0pt plus 1.0pt minus 2.0pt   
\def\href#1#2{#2}	

\def\coeff#1#2{{\textstyle {\frac {#1}{#2}}}}
\def\half{\coeff 12}

\def\N{{\cal N}}

\def\R{{\mathbb R}}
\def\S{{\mathbb S}}

\def\tr{{\rm tr}}

\def\Z{{\mathbb Z}}

\def\Dslash{{\rlap{\raise 1pt \hbox{$\>/$}}D}}

\def\SN{\mathcal S_{N}}

\title{ 
Seiberg-Witten   and ``Polyakov-like"  magnetic bion   confinements     are continuously connected 
}

\author
    {
    {
    \def\href#1#2{#2}	
    Erich Poppitz$^1$\footnote{\email{poppitz@physics.utoronto.ca}}~
    and Mithat \"Unsal$^2$\footnote{\email{unsal@slac.stanford.edu}}~
           \\${}^1${Department of Physics, University of Toronto,
    Toronto, ON M5S 1A7, Canada}
     \\${}^2${SLAC and Physics Department, Stanford University, Stanford, CA 94025/94305, USA}
        }
    }%

\abstract{

\smallskip

{\small{
We study 
four-dimensional $\N=2$ supersymmetric  pure-gauge  (Seiberg-Witten)  theory and its  $\N=1$  mass perturbation   by using compactification  on  $\S^1 \times {\mathbb R}^{3}$.  It is well known that on $\R^4$ (or at large $\S^1$ size $L$) the perturbed   theory realizes confinement through monopole or dyon  condensation.   At small  $\S^1$, we demonstrate that   confinement is induced by a  generalization of Polyakov's  three-dimensional instanton mechanism    to a locally four-dimensional theory---the  magnetic bion mechanism---which also applies to a large class of nonsupersymmetric theories. Using a  large- vs.  small-$L$ Poisson duality, we show that  the two mechanisms of confinement, 
  previously thought to be distinct,   are  in fact  continuously connected. 
 }}}
 
\begin{document}

\maketitle

 \section{Introduction and results}

 In this paper, we study the dynamics of $\N=2$ supersymmetric pure  gauge (Seiberg-Witten \cite{Seiberg:1994rs}) theory  and its  $\N=1$  mass perturbation  compactified        on   $\R^3 \times \S^1$ through a new method.  
  We mostly work with an $SU(2)$ gauge group and only mention  $SU(N)$ in connection with   non-'t Hooftian (abelian) large-$N$ limits.   
Ref.~\cite{Seiberg:1996nz} already examined  this  theory   on  $\R^3 \times \S^1$. A   description of the  vacuum structure of the theory is given as a function of the circle radius $L$, interpolating between 3d and 4d results.    Supersymmetry, holomorphy, and  elliptic curves provide much information about the vacuum of the theory.  However,  many physical aspects  of the mass-perturbed  $\N=2$ theory   on $\R^3 \times \S^1$ remain open.  
 For example, at small $\R^3 \times \S^1$,  one can ask:
  \begin{itemize}
 \item[{\it i)}] what generates confinement and the mass gap for gauge fluctuations?
   \item[{\it ii)}] what   induces chiral symmetry breaking and generates mass for fermions?
\item[{\it iii)}]  what  stabilizes the center symmetry? 
  \end{itemize}
  These are questions of interests not only in the supersymmetric theory, but also of central importance in  non-supersymmetric   QCD and QCD-like gauge theories on  $\R^3 \times \S^1$.     It turns out that  adequately  answering these questions opens interesting 
 avenues in the study of confinement and topological defects  in gauge theories, not exclusively restricted to supersymmetric theories.

 \subsection{Method } 
 
\begin{figure}[h]
 \begin{center}
\includegraphics[angle=-90, width=3.0in]{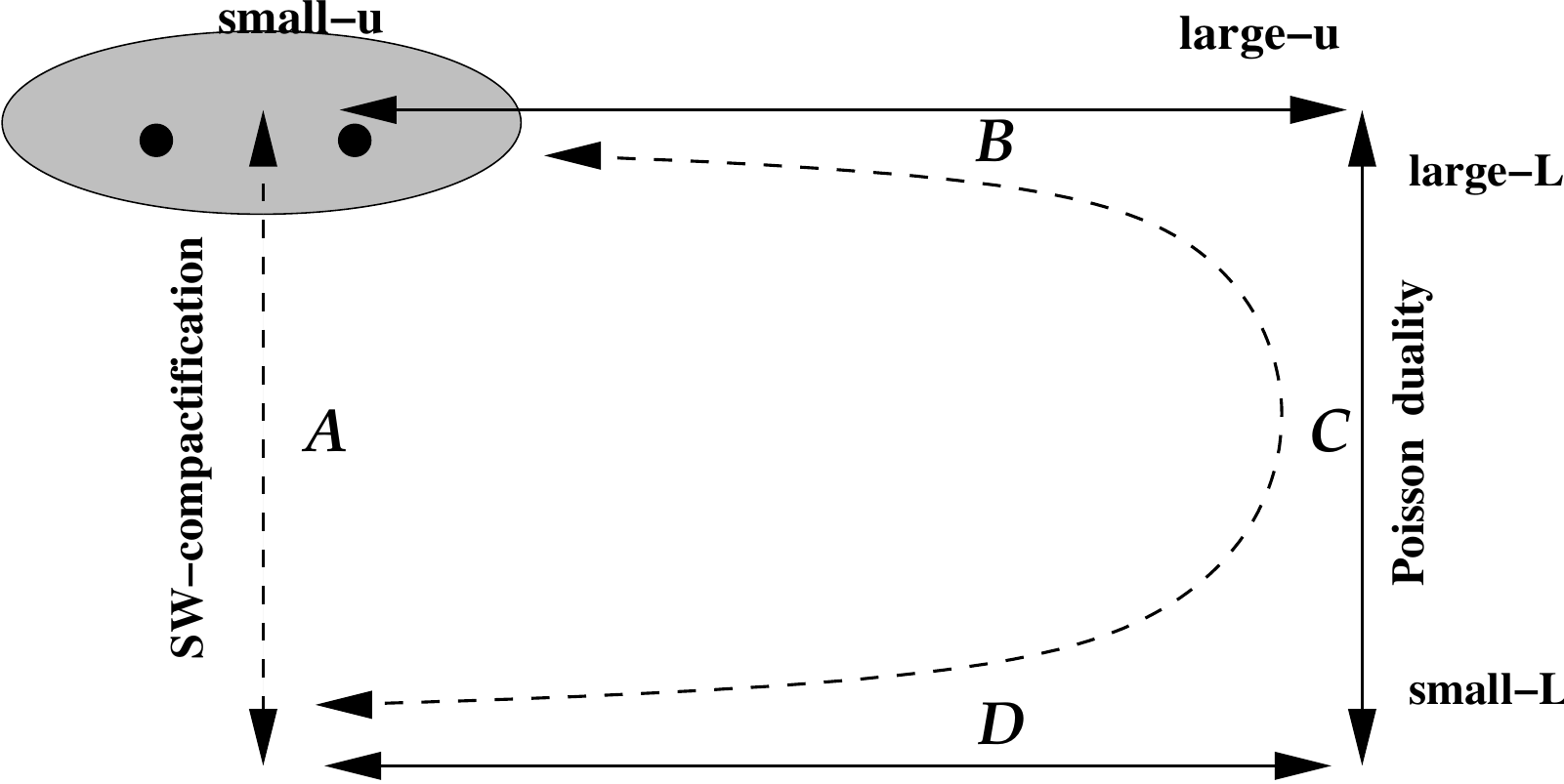}
\caption{ 
Taking different paths in the $u$-$L$ plane. The horizontal direction, $u$,  is the modulus of Seiberg-Witten theory and the vertical, $L$,  is the size of $\S^1$.
Ref.~\cite{Seiberg:1996nz} studied the softly broken $\N=2$ theory on $\R^3 \times \S^1$  by using 
 elliptic curves through path ${A}$.  In this work, we reexamine the same theory along the path $BCD$ in moduli space. The $CD$ branch   always remains semi-classical and allows us to understand the relation between the topological defects responsible for confinement at small-$L$ and large-$L$ in detail.  
 }
  \label {fig:ourpath}
 \end{center}
 \end{figure}

In this work, we use a  different methodology relative to  Ref.~\cite{Seiberg:1996nz} to study the theory on    
 $\R^3 \times \S^1$.  Our approach, shown in  the commutative diagram in Fig.~\ref{fig:ourpath},  permits us to study this theory by using simple field theory techniques. Some of our techniques also   apply to non-supersymmetric theories. 
 
  Let us now briefly describe the physics of Seiberg-Witten solution at $\R^4$ and the reasoning behind  Fig.~\ref{fig:ourpath}. 
 The $\N=2$ theory on $\R^4$ possesses a quantum moduli 
  space parameterized by $u=  \langle \tr \Phi^2 \rangle$,  where $\Phi$ is an adjoint chiral multiplet.    The $u$-modulus also provides a control parameter. On the moduli space, the $SU(2)$  gauge symmetry is Higgsed down to  $U(1)$   at a scale $|u|^{1\over 2}$. Since the theory is asymptotically free,   for $|u| \gg \Lambda_{\N=2} ^2$, where $\Lambda_{\N=2} $ is the strong scale,   the theory is (electrically) weakly coupled, $g_4^2 (|u|) \ll 1$.  
  (From now on, we set $\Lambda_{\N=2} =1$).  
   The $|u| \lesssim 1$ domain (shaded region in Fig.~\ref{fig:ourpath}) is electrically strongly coupled.  
   
      The $SU(2)$$\rightarrow$$U(1)$ theory possesses the 't Hooft-Polyakov monopole and dyon particles, which are heavy at $|u| \gg 1$. 
 There are two points on the
moduli space  in the shaded region  in Fig.~\ref{fig:ourpath},  where a monopole ($u= +1$)  or a dyon ($u= -1$)  become massless. The low-energy limit of the theory near
the monopole (or dyon) points  is described by the  $\N=2$  supersymmetric electrodynamics (SQED)  of massless monopoles (or dyons).  The gauge field and the coupling in SQED are   dual  to the ones in the microscopic theory.    In particular, whenever the electric coupling is large, the dual magnetic coupling is small and vice versa. The effective field theory descriptions near the  $u= +1$ and $u=-1$ points  are mutually non-local and there is no global macroscopic theory which describes both  $u=+1$ and  $u=-1$.  Physically, one of the most interesting outcomes of the Seiberg-Witten solution  is that   when the $\N=2$ theory is perturbed by an $\N=1$ preserving mass term for $\Phi$, it exhibits confinement of electric charges due to magnetic monopole or dyon condensation.

 Ref.\cite{Seiberg:1996nz} studied the  $\N=2$  SYM and its softly broken  $\N=1$ version   on $\R^3 \times \S^1$  by using   elliptic curves through path-${A}$ on Figure~\ref{fig:ourpath}. However, if we would like to understand the relation between the topological defects  (and field theories) at large   and small $\S^1$,  there are some intrinsic difficulties associated with path-${A}$. 
 In particular, the large-$\S^1$ theory is magnetically weakly  and electrically strongly  coupled, and the small-$\S^1$ one is electrically weakly (by asymptotic freedom)  and magnetically strongly coupled. Thus, when $L \sim 1$ and $|u| \lesssim  1$, both electric and magnetic couplings are order one, and we do not know how to address this domain in field theory.  
 To avoid this difficulty, we propose  a compactification (path-${C}$)  at large-$u$ where the theory is always electrically weakly coupled, regardless of the $\S^1$-size $L$.  Path-${D}$ is also always   weakly coupled, either because the $u$-modulus is large or because an additional modulus, the Wilson line along $\S^1$, is turned on (also note that in the small-$L$ domain the $\N=2$ theory always abelianizes, and the long-distance dynamics is described by a three dimensional hyper-K\" ahler nonlinear sigma model \cite{Seiberg:1996nz}).
  
\subsection{Conclusions}

    We find, by using  the techniques of Ref.\cite{Seiberg:1996nz} and  of our current work,  that  a  locally  four dimensional generalization \cite{Unsal:2007vu, Unsal:2007jx}  of Polyakov's 3d instanton mechanism of confinement    \cite{Polyakov:1976fu} takes over in the small-$L$ mass-perturbed $\N=2$ theory. 
To elucidate, note that  the theory possesses 3d instanton (and anti-instanton) solutions, which, when embedded in $\R^3 \times \S^1$, have  magnetic, $Q_m=\int_{\S^2_\infty} F$,  and topological, $Q_T=\int_{\R^3 \times \S^1} F \widetilde F$ charges, normalized  to  $ (Q_m,Q_T)= \pm \left(1, \half\right)$. There are also 
twisted instantons  (and anti-instantons), which carry charges  $(Q_m,Q_T)= \pm \left(-1, \half\right)$. The  mass gap for gauge fluctuations  and confinement in  the mass-perturbed $\N=2$ theory 
arise due to Debye screening by topological defects  with charges  $(Q_m,Q_T)=\left(\pm 2, 0\right)$.   This mechanism of  confinement was called the ``magnetic bion"  mechanism in  \cite{Unsal:2007vu, Unsal:2007jx} and we show here that it also takes place in the  $\N=1$ mass deformation of Seiberg-Witten theory at small $L$.   The fact that the leading  instanton amplitude  in the  semi-classical expansion  cannot generate mass gap for gauge fluctuation---which distinguishes the  magnetic bion mechanism from  Polyakov's  3d instanton mechanism---is due to the presence of fermion zero modes, dictated by the Nye-Singer index theorem for the Dirac operator on $\S^1 \times \R^3$ \cite{Nye:2000eg, Poppitz:2008hr}.

 {\flushleft{O}}ur main conclusions are: 
 \begin{enumerate}
 \item There are two types of confinement mechanisms in mass-perturbed $\N=2$ theory. At $L$ large compared to the inverse strong scale of the theory, confinement is due to  magnetic monopole or dyon condensation. At small $L$, it is the ``Polyakov-like"    magnetic bion mechanism briefly described above.
 \item Under the reasonable  assumption  that  supersymmetric theories with supersymmetry-preserving boundary conditions on $\S^1 \times \R^3$ do not have any phase transition as a function of radius, these two mechanisms ought to be continuously connected. The physical questions   we address in this work are: 
\begin{enumerate}
\item What is the relation between the monopole and dyon particles  on  $\R^4$  (or  large  $ \S^1 \times \R^3$) and the monopole-instantons and magnetic bion-instantons of the small $\S^1 \times \R^3$\
regime?  How do we relate the two confinement mechanisms? 
\item   What is  the region of validity  of the various small- and large-$L$  descriptions? 
\end{enumerate}
Our results show that the relation between the topological defects responsible for confinement at small $L$ and large $L$ is   intricate---even in the case where confinement remains manifestly abelian at any $L$, as in the  mass-perturbed Seiberg-Witten theory. However, along the path-${C}$ in   Fig.~\ref{fig:ourpath} of undeformed theory, we find a 
precise duality relation between the semi-classical topological defects pertinent to confinement at large- and small-$L$.  
More precisely, the 3d monopole-instantons and twisted monopole-instantons, which make up the magnetic bion ``molecules" that generate mass gap and confinement at small $L$,  have a Kaluza-Klein tower. The nonperturbative contribution of this tower is   dual, through a Poisson resummation,   to that of the tower of 4d monopole/dyon particles whose Euclidean worldlines wrap around the compact direction. We refer to the duality along the
 path-${C}$ as {\it Poisson  duality}. This duality presents an explicit relation between the topological defects responsible for confinement at small $\R^3 \times \S^1$ and on $\R^4$. 
 
  \item The magnetic bion mechanism also holds in a large-class of non-supersymmetric theories at small $L$. Thus, 
 our construction gives a map between  theoretically controllable  confinement mechanisms in non-supersymmetric and supersymmetric gauge theories. 
  \end{enumerate}

\subsection{Outline}

We begin in Section~\ref{sec:classical} by reviewing the classical pure $\N=2$ supersymmetric Yang-Mills theory, both using 4d notation (Section~\ref{sec:review}) and dimensional reduction from 6d (Section~\ref{sec:6d}). The latter is useful when studying the supersymmetries preserved by various classical solutions in Appendix~\ref{chiralityappendix}. In Section~\ref{sec:6d}, we also introduce come useful notation.

We begin the discussion of the classical solutions in  Section~\ref{sec:relating} by recalling, in Section~\ref{sec:mondyon},   the properties of  monopole and dyon particles on $\R^4$. The corresponding tower of  monopole- and dyon-instantons on $\R^3\times \S^1$, pertinent to the large-$L$ nonperturbative dynamics, is constructed in Section~\ref{sec:mondyonatlargeL}. 
In Section \ref{sec:bos}, we describe the  tower of winding monopole-instanton solutions  at a generic point in moduli space, relevant to the small-$L$ dynamics.
The Poisson duality between the 3d tower of winding solutions and 4d tower of dyon-instantons is discussed in the following three Sections. In Section \ref{poissondualitysection}, the duality is discussed and qualitatively explained in a simplifying limit.
In Section \ref{poissongeneral}, a more general duality relation is derived and then discussed in Section \ref{digression}.

In Section~\ref{massdeformed}, we study the role the winding monopole-instantons and dyon-instantons, discussed above, play in the nonperturbative dynamics of confinement and chiral symmetry breaking at large or small $L$.
 In Section  \ref{largeLSW}, we recall the SW description of monopole/dyon condensation in the mass-perturbed theory and give the large-$L$ expressions for the mass gap and string tension. Then, we explain the difficulties a compactification along path A of Fig.~\ref{fig:ourpath} would face. We also give an effective 3d description of the physics at scales larger than $L$, valid for $L\Lambda_{\N=2} \gg 1$, using a chain of known 3d dualities.

 The small-$L$ dynamics is studied in Section~\ref{sec:smallLdynamics}, beginning with a discussion of the 't Hooft vertices induced by the winding monopole-instantons of lowest action. Then, in Section~\ref{sec:massatsmallL}, we explain  the effect of  the $\N=1$ preserving mass perturbation, the generation of a superpotential, the resulting vacua of the theory, and give expressions for the mass gap and string tension (all results well-known in the literature). Then, we concentrate on the physical mechanisms responsible for the center-symmetry stabilization and confinement. We note that they are due to different kinds of instanton--anti-instanton ``molecules." 
Most notably, we explain that the mass of the dual photon is generated by magnetic bions, bound states of monopoles and twisted anti-monopoles of magnetic charge two, and an example of a topological ``molecule" whose stability is semiclassically calculable.
 The realization of the unbroken center-symmetry is discussed in Section~\ref{sec:centersymmetry}. The physics of chiral symmetry breaking is discussed in Section~\ref{sec:chiralsymmetry}, along with an elaboration on some imprecise statements in the literature.  
  
In Section~\ref{phase}, we   discuss  the phase diagram in the $m$-$L$ plane, indicating the regimes where the different topological excitations discussed above play a role in the confinement mechanism. 
We consider  
  both the $SU(2)$ (small-$N$) and 
 large-$N$ cases. In the latter case, we note that the abelian description 
 of the dynamics persists as $N \rightarrow \infty$ only in non 't Hooftian large-
$N$ limits, both at small $L$ (where $L$ must scale as $1/N$) and large $L$ (where the soft breaking mass $m \sim 1/N^4$).
 
We summarize our findings  and discuss some open problems in Section \ref{conclusions}. 
 We give various technical details in the appendices. In Appendix~\ref{chiralityappendix}, we study the supersymmetries preserved by the various solutions discussed in the paper, in order to identify the nature of the  unlifted supersymmetric fermion zero modes. 
In Appendix~\ref{sungeneral}, we generalize to $SU(N)$ the  Poisson duality relation  of Section~\ref{poissongeneral}. 
 
 \section{Review of the classical  $\mathbf{ \N=2}$ supersymmetric theory } 
 \label{sec:classical}
 
 \subsection{Theory on $\mathbf{\R^4}$  and global symmetries}
 \label{sec:review}
The matter content and bosonic symmetries of the pure   $\N=2$  supersymmetric Yang-Mills theory on $\R^4$ fill  a representation of  the $ SO(4) \sim [SU(2)_{\cal L}  \times SU(2)_{\cal R} ]_E$ Euclidean Lorentz symmetry and the $SU(2)_{R} \times U(1)_{R}$ chiral R-symmetry.    The transformation properties under 
$[SU(2)_{\cal L}  \times SU(2)_{\cal R}  \times SU(2)_{R} ]_{U(1)_{R}}$ are as follows:  the gauge field  $A_{\mu} \sim  (\half, \half, 0)_0$, the scalar 
  $\phi \sim (0, 0, 0)_{+2}$, fermions  $\lambda^i   \sim  (\half, 0 ,\half)_{+1}$.
  All fields are valued in the adjoint representation of the gauge group  $G$ and fill a gauge multiplet  of  $\N=2$ supersymmetry.  
   In this paper, we study mostly $G=SU(2)$ and  give a generalization to 
  $SU(N)$ for some of the results.  

The  $\N=2$ supersymmetric  gauge multiplet  diamond  can be decomposed in terms of   $\N=1$  multiplets,  vector 
$V= (A_\mu, \lambda)$ and chiral  $\Phi= (\phi, \psi)$ multiplets, as well as  
$\N=1'$  $V'= (A_\mu,  \psi) $,  $\Phi'= (\phi,  \lambda)$ multiplets 
as  shown below:
\begin{equation}
 \xymatrix{
& A_{\mu} \ar@{<->}[dl]_{\N=1}  \ar@{<.>}[dr]^{\N=1'}  & \\ 
 \lambda  \ar@{<.>}[dr]_{\N=1'}  &  &   \psi \ar@{<->}[dl]^{\N=1}    \\
& \phi &
 }  
\end{equation}
We will eventually be interested in the theory with only $\N=1$ supersymmetry, 
where ${\N=1'}$ part of the supersymmetry is broken by a soft mass term for the $\Phi$-multiplet. 
We parameterize the $SU(2)_R$ doublet  as $ \left( \begin{array}{c} \lambda^1 \\ \lambda^2 \end{array} \right) = \left( \begin{array}{c} \lambda \\ \psi \end{array} \right)  $, where the first form is used whenever we want to make $SU(2)_R$ invariance manifest. 

The Lagrangian of $\N=2$ Yang-Mills  theory  may be written in component fields as: 
\begin{equation} \label{action2}
{\cal L} = \frac{2}{g_4^2} \tr \left[ \frac{1}{4} F_{\mu \nu}^2 + D_{\mu} \phi^{\dagger}D_{\mu} \phi +  \half [  \phi^{\dagger},  \phi  ]^2       
 + i \overline \lambda_i   \overline \sigma_{\mu} D_{\mu}\lambda^i     - \frac{i}{\sqrt 2} \epsilon_{ij}  \lambda^i  [\lambda^j, \phi^{\dagger}]    - \frac{i}{\sqrt 2} \epsilon^{ij}    \overline  \lambda_i    [\overline \lambda_j , \phi]  \right]~,
\end{equation}
where $D_{\mu}  = \partial_{\mu}  + i [ A_{\mu}, \; ] $ and the field strength is
$F_{\mu \nu}=  \partial_{\mu}  A_{\nu} -   \partial_{\nu}  A_{\mu} + i   [A_{\mu},   A_{\nu} ]$. We normalize the Lie algebra generators as $\tr \;t^a t^b= \half \delta^{ab}$. 
The component formalism makes the chiral symmetries manifest, but it hides the exact $\N=2$
supersymmetry. 

The classical    $U(1)_R$ symmetry is anomalous quantum mechanically. For an $SU(N)$ gauge group, 4d instantons generate 
an amplitude:
 \begin{equation}
 e^{-S_I}  \left[ \half \epsilon_{i_1i_2}  \epsilon_{j_1j_2}  (\lambda^{i_1} \lambda^{j_1})  (\lambda^{i_2} \lambda^{j_2})   \right]^N  \equiv e^{-S_I} [\det_{i,j} (\lambda^i \lambda^j)]^N~,
 \label{4d-instanton}
   \end{equation}
which is manifestly invariant  under $SU(2)_{\cal R}$, but rotates by a phase 
$e^{ i\;   4 N \alpha  }$ under  $U(1)_R$, $\lambda^i \rightarrow e^{i \alpha} \lambda^i$. 
Thus, the quantum theory  respects only a    $\Z_{4N}$ subgroup of  $U(1)_R$. 
For $SU(2)$ gauge group,  the exact chiral symmetry of the quantum theory is:
\begin{equation}
(SU(2)_R \times \Z_{8}) /\Z_2
\end{equation}
 where $\Z_2$ is  factored out to prevent double counting of the factor $(-1)^F$ (where $F$ is fermion number) common to the center of $SU(2)_R$ and  $\Z_{8}$. 

Since the $\N=2$ Lagrangian (\ref{action2}) includes  terms of the form $\phi^{\dagger} [\lambda_1, \lambda_2]$, the scalar $\phi$ is also charged under $\Z_{8}$. It transforms as 
  $\phi \rightarrow  e^{i {\pi\over 2}}\phi$. The simplest gauge invariant that we build out of 
  $\phi$ is $u\equiv \tr \phi^2$.  The $u$ field parametrizes the classical moduli space of gauge theory, and   
   it changes sign under the    $\Z_{8}$ action.  This discrete symmetry will be crucial once we consider the theory on $\R^3 \times \S^1$. To summarize, 
 the  action of the anomaly-free chiral   $\Z_{8}$  symmetry is: 
   \begin{equation}
 \Z_8: \lambda^{i} \rightarrow  e^{i {2\pi\over 8}}\lambda^i, \qquad \phi  \rightarrow  e^{i {4\pi\over 8}} \phi,
    \qquad u \rightarrow -u .
  \end{equation}
  Note that  the  $\Z_8$ symmetry is unbroken only at the $u=0$ point in the classical moduli space.  This will also have interesting consequences for the   theory on $\R^3  \times \S^1$.

\subsection{Reduction of six dimensional $\mathbf{\N=1}$ theory and notation}
\label{sec:6d}
It will be  also useful to describe the $\N=2$ theory in 4d   by starting 
with the  minimal supersymmetric Yang-Mills theory in 6d  with the Lagrangian:
\begin{eqnarray}
{\cal L} && = \frac{2}{g_4^2} \tr \left[ \frac{1}{4} F_{MN}^2 +  i \overline \Psi   \Gamma_M  D_M \Psi      \right]  \qquad
 \label{6d}
\end{eqnarray}
where $M, N=1, \ldots, 6$.  $\Gamma^{M}, M=1, \ldots 6$ denote the six  gamma-matrices satisfying the Clifford algebra $\{\Gamma^{M}, \Gamma^{N}\} = 2 \delta_{MN}$, and $\Gamma^{7}$ is the chirality matrix in six dimensions.\footnote{ We may use the following basis for computations:
\begin{equation}
\Gamma^{\mu}=  \sigma_2 \otimes \gamma_\mu, \qquad \Gamma^{5}=  \sigma_2 \otimes \gamma_5, \qquad  \Gamma^{6}=  \sigma_1 \otimes 1_4, \qquad  \Gamma^{7}=  \sigma_3 \otimes 1_4,
\end{equation}
where  $\gamma_{1,\ldots, 4} $ are four-dimensional gamma matrices and $\gamma_5= \gamma_1\ldots \gamma_4$.  However, using an explicit basis is not necessary for our purposes, see Appendix \ref{chiralityappendix}.} The complex spinor $\Psi$ satisfies the chirality condition, $(\Gamma_7 +1) \Psi=0$.

The   $\N=1$  theory in 6d has an $SU(2)_R$ chiral symmetry which acts on fermions (it is not manifest in the way we have written it).  
The fermions as well as the supercharges transform as doublets under this symmetry, whereas the gauge field is a singlet. 
Dimensional reduction turns the  Lorentz symmetry in the reduced directions into global $R$-symmetries of the lower dimensional  theory. Let us denote the Euclidean spacetime directions as $x^{1,2,3,4,5,6}$.   The $\N=2$ theory on $\R^4$ may be obtained  by erasing the $x^{5,6}$ dependence from all fields.  This means that the  $SO(6)_E$  Euclidean Lorentz symmetry transmutes to $SO(4)_E \times SO(2)_R$ symmetry, 
 whose covering group  is  $[SU(2) \times SU(2)]_E \times U(1)_R$.   Together with the $SU(2)_R$ mentioned above, this is the symmetry group of the 4d $\N=2$ theory described in  Section~\ref{sec:review}.
 
 Let us denote the gauge  field of the six dimensional theory as  $A^{M}$.  Consider both dimensional  reduction and compactification\footnote{We distinguish dimensional reduction and compactification. 
Compactification,  unlike dimensional reduction,   does not alter the microscopic  chiral symmetries of the theory, which  has important consequences.} down to  to   $\R^4$, $\R^3 \times \S^1$, and $\R^3$.   
The 6d gauge field decomposes as follows:
\begin{eqnarray}
\label{notation431}
&&A_M \rightarrow  A_{\mu} \oplus  
\underbrace{A_5,  A_6}_{ \phi,  \; \phi^\dagger},  \;\;  \mu=1, \ldots, 4, \qquad \R^4  \cr
&&A_M \rightarrow  A_{i} \oplus \underbrace{A_4}_{b}  \oplus    
\underbrace{A_5,  A_6}_{ \phi,  \; \phi^\dagger},  \;\; i=1,2, 3, \qquad \R^3 \times \S^1
   \end{eqnarray}
where $ \phi= (A_5 + i A_6)/\sqrt 2$.  If the gauge theory abelianizes, at scales larger than the $\S^1$ size $L$  we can dualize the three dimensional field strength to a compact scalar $\sigma$, via: 
\begin{equation}
F_{ij}  =   {g_4^2 \over 4 \pi L} \epsilon_{ijk} \partial_k \sigma \;, 
\label{dual}
\end{equation}
and use the complex fields $B= (b+ i\sigma)/{\sqrt 2}$ to obtain the decomposition:
\begin{eqnarray}
\label{notation3}
&&A_M \rightarrow  \underbrace{ \sigma, b}_{B, B^{\dagger}}  \oplus    
\underbrace{A_5,  A_6}_{ \phi,  \; \phi^\dagger},  \qquad  b \equiv \frac{4 \pi}{g_4^2}\omega, ~ \omega \equiv L A_4 \qquad \R^3~.
   \end{eqnarray}
In (\ref{notation3}) $A_4$ refers to the value of $A_4$ in the Cartan subalgebra (the unbroken abelian gauge group) and $\omega$ denotes the corresponding Wilson line around the compact direction.

 \section{Relating 4d monopole particles to 3d monopole-instantons}
 \label{sec:relating}
 
 \subsection{Monopole and dyon particles on $\mathbf{\R^{4}}$ }
 \label{sec:mondyon}
 
 Consider the pure $\N=2$ theory in the semi-classical domain of its moduli space. Denoting $v =  |u|^{1\over 2}$, this is the regime where $v \gg 1$; recall that we set $\Lambda_{{\cal{N}}=2}$$=$$1$.  
 BPS particles  with electric and magnetic charges  $(n_m, n_e)$ have masses determined by the central charge $Z_{(n_m, n_e)}= v(n_e + n_m \tau)$, where  $\tau = {4 \pi i \over g_4^2} + {\theta \over 2 \pi}$ is the holomorphic gauge coupling: 
 \begin{eqnarray}
  \label{bpsmass1}
M(n_m, n_e) = |Z_{(n_m, n_e)}| = v \sqrt {  n_m^2 \left( \frac{4 \pi}{g_4^2} \right)^2 + \left( n_e + n_m \frac{\theta}{2\pi } \right)^2 }  =  v \sqrt {  n_m^2 \left( \frac{4 \pi}{g_4^2} \right)^2 + n_e^2 } ~.
 \end{eqnarray}
 From now on, we set $\theta=0$. 
 
\begin{figure}[t]
 \begin{center}
\includegraphics[angle=-90, width=4.5in]{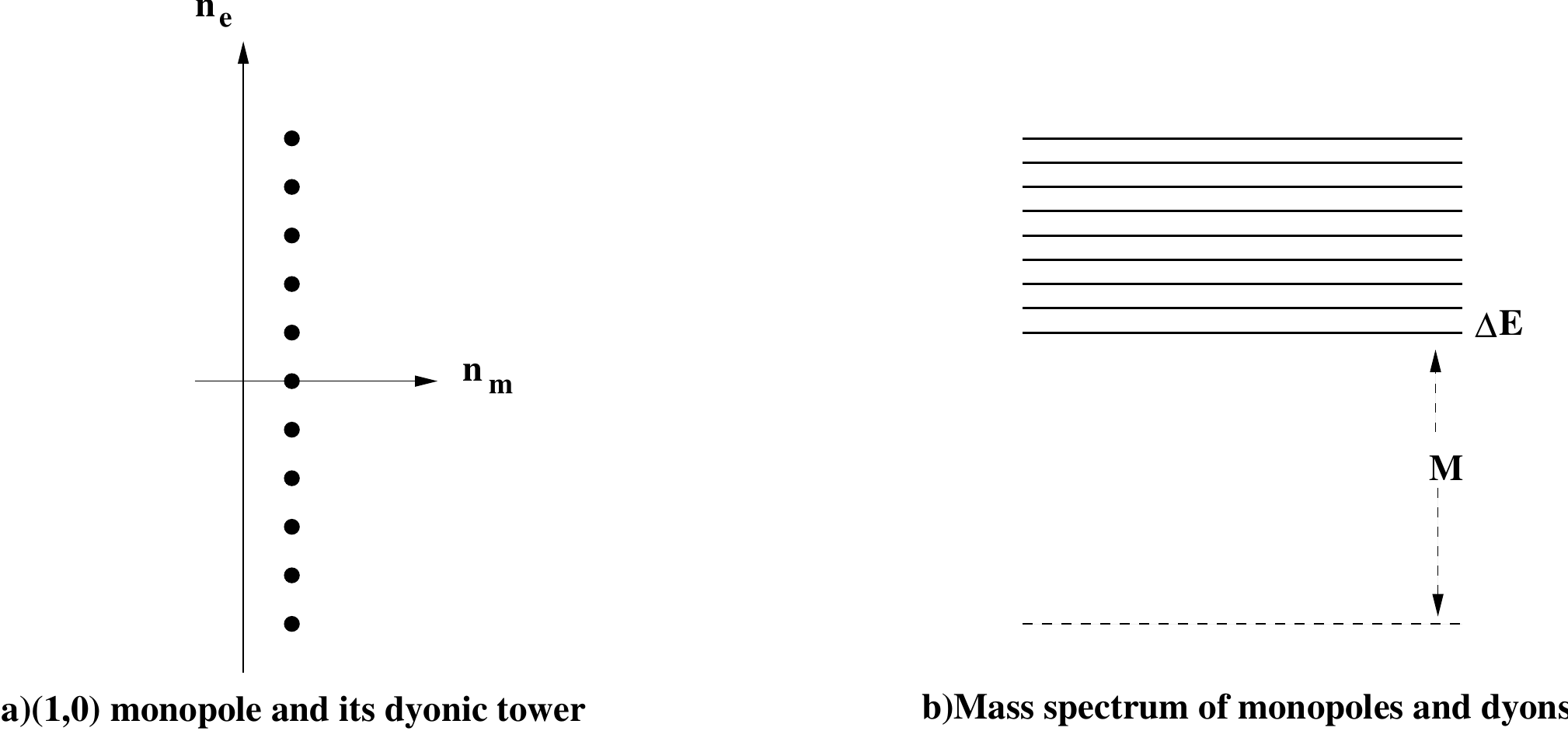}
\caption{a)The spectrum of charges of a monopole and its dyonic tower obtained by quantizing 
the $U(1)_e$ zero mode.  b) The mass spectrum. In the semi-classical regime, $\Delta E\equiv  
E_{(1, \pm 1)} - E_{(1,0)}   \ll M$.  This tower of  states, labelled by electric charge, is  pertinent to large-$L$ and is  Poisson-dual to the 3d BPS monopole-instanton and its tower, characterized  by the  winding number, and pertinent to small-$L$. }
  \label {fig:dt}
 \end{center}
 \end{figure}

 In the limit $v \gg 1$, the coupling is small $g_4^2 (v) \ll1$. The monopole with $n_m=1$  and arbitrary $n_e$ will be relevant below.  Its mass is given by expanding (\ref{bpsmass1}): 
   \begin{eqnarray}
   \label{bpsmass2}
M(1, n_e) \approx  \frac{4 \pi v }{g_4^2}  + \frac{1}{2}  \frac{g_4^2 v}{4\pi} n_e ^2  =  M_{(1,0)} +  ( M_{(1,n_e)} -  M_{(1,0)} ) \equiv  M_{(1,0)} + \Delta M_{(1,n_e)}~.
 \end{eqnarray}
This formula has a well-known  physical interpretation. A  monopole in the four-dimensional theory has four collective coordinates. The classical solution is not invariant under  three spatial translations and under the unbroken subgroup 
 $U(1)_e \subset SU(2)$. The corresponding collective coordinate space is  $ ({\vec a}, \varphi)  \in  \R^3 \times \S^1_\varphi$. The angular zero mode  is generated by $U(1)_e$ rotations and the eigenvalue of rotation on the unit circle $\S^1_\varphi$ is the electric charge. 
 The wave function associated with the collective coordinates is $\Psi  ({\vec a}, \varphi )  = e^{i \vec p \cdot  \vec a} e^{i n_e \varphi}$.   
 When we quantize the BPS monopole with $n_m=1$,  we observe that it can carry arbitrary electric charge, an  integer multiple   of the fundamental charge,  as shown in Fig.~\ref{fig:dt}a (such  towers exist for any magnetic charge monopole $(n_m,0)$, albeit they may  be unstable). The dyonic tower of the anti-monopole is  $(-1, n_e), \; n_e \in \Z$. 
 
 In the semi-classical regime,  the mass  of a $(1, 1)$ dyon (\ref{bpsmass2})  differs negligibly from that  of a monopole.   In fact, a large number of states occupying the dense band shown in  Fig.~\ref{fig:dt} (clearly, the spectrum is not equidistant as may appear from the figure) has
$ \Delta M_{(1,n_e)} \ll M_{(1,0)}$. The  states almost degenerate with the 
monopole have $|n_e| \ll n_e^{\rm max} \ll {4\pi}/g_4^2$.  When $|n_e| < n_e^{\rm max}$, the fermionic zero modes  of the states in the tower will also be identical, at leading order; this will be important for our future considerations.

Also for future reference we note that the  Bogomolnyi's bound  applied to  $(1, n_e)$ BPS monopole/dyon particles yield the first order differential equations (see Ref.~\cite{Weinberg:2006rq} for a review):
   \begin{eqnarray}
 &&  \vec B  - \cos \delta_{n_e}  \vec D A_5 =0 ~,\cr
 &&  \vec E - \sin \delta_{n_e}   \vec D A_5 =0 ~,\cr
 &&D_4A_5=0 ~,
   \label{delta}
 \end{eqnarray}
 where:
 \begin{equation}
e^{i \delta_{n_e}} =   \frac{ \frac{4 \pi}{g_4^2}  + i n_e } {
\sqrt {   \left( \frac{4 \pi}{g_4^2} \right)^2 + n_e^2 } }  ~.
\label{cosdelta}
 \end{equation}
 We have rotated the scalar  vev, see (\ref{notation431}), in the $\phi = (A_5, A_6)$ plane to purely $A_5$, with no loss of generality.

\subsection{Monopole-instantons and dyon-instantons at  large $\mathbf{\S^1 \times \R^3}$} 
\label{sec:mondyonatlargeL}

To study the theory on $\S^1 \times \R^3$, we compactify the $x_4$ direction on a circle with circumference $L$. When $L \Lambda_{\N=2} \gg 1$, the spectrum of the theory is clearly that of Seiberg-Witten theory on $\R^4$ with trivial restrictions due to the boundary condition in $x_4$. The perturbative spectrum consists of photons, electrically charged $W$-bosons, and their superpartners, while the non-perturbative spectrum is comprised of  monopoles, dyons, and their superpartners. 

In the $v  \gg 1 $ regime, the nonperturbative magnetically charged states are semiclassically accessible and are thus significantly heavier then the perturbative states. They do, however, contribute to the dynamics of the quantum theory on $\S^1 \times \R^3$. If one lets the Euclidean worldline of a monopole/dyon particle  wrap around the $\S^1$, this ``pseudo-particle"   acquires a finite Euclidean action,   
$S(n_m, n_e)= LM(n_m, n_e)$. This  
 means that it has  to be interpreted not as a (BPS) state in the compactified  theory, but rather as an instanton of action, which, for $n_m=1$, is given by:\footnote{The  theory also has 4d instantons, obeying the self-duality condition $F_{\mu \nu}= \half \epsilon_{\mu \nu \rho \sigma} F^{\rho \sigma}$ of action $\frac{8 \pi^2}{g_4^2}$. As these carry no magnetic charge, they are not  relevant for  confinement at small $L$.}
    \begin{eqnarray}
    \label{dyonaction}
S(1, n_e)= LM(1, n_e)  =vL  \sqrt {  \left( \frac{4 \pi}{g_4^2} \right)^2 +  n_e^2 }  \; .
   \end{eqnarray}
These instantons represent saddle points of the Euclidean path integral and their contributions must be summed over. 
  
  When we consider the theory on $\R^4$, we can gauge away the gauge field in any one chosen direction, in particular its $x_4$-component $A_4$. However, once the theory is compactified, $x_4$$\equiv$$x_4 + L$, the zero mode of $A_4$---equivalently, the Wilson line around $\S^1$---can no longer be gauged away. In the supersymmetric theory on $\S^1 \times \R^3$, we are free 
   to turn on an arbitrary constant and homogeneous $A_4$ background gauge field
    commuting 
       with the vev of $\phi$, $A_4 = a_4 T^3$. This background gauge field naturally couples to the electric charge of the dyonic tower, modifying the action:
   \begin{equation}
   \label{dyonaction2}
   S(1, n_e) \rightarrow S(1,n_e) + i n_e \int\limits_0^L d x^4 a_4 = S(1, n_e) + i n_e a_4L
   = S(1, n_e) + i n_e \omega~,
   \end{equation}
   where in the last line we 
    recalled the definition (\ref{notation3}) of $\omega$, an angular variable (see the following Section~\ref{sec:bos}).

The   ``dyon-instantons"  with charges $(1, n_e)$  induce amplitudes 
$\sim e^{ - S(1,n_e)}$ $e^{i  \sigma +  i n_e \omega  }$ $\times$ $({\rm fermion\;  
zero \; modes})$
in the long distance effective Lagrangian, where $\sigma$ is the dual photon 
defined in (\ref{dual}). (The fermionic zero modes are discussed in Sections 
\ref{poissondualitysection}, \ref{poissongeneral}, and Appendix \ref{chiralityappendix}.)
 The $e^{i \sigma}$ factor, within the dilute gas approximation,  takes into account the long-distance Coulomb interactions between dyon-instantons \cite{Polyakov:1976fu}. 
    Thus,  in the large-$L$  limit, the sum of the leading semiclassical contributions 
   with magnetic charge $n_m =1$  comes from the  infinite dyonic tower  $(1, n_e), n_e \in \Z$.   
   This gives, schematically:
    \begin{eqnarray}
({\rm Dyon \;sum \;at \;large-}L) \sim e^{i \sigma} \sum_{n_e \in \Z} e^{ - vL  \sqrt {  \left( \frac{4 \pi}{g_4^2} \right)^2 +  n_e^2 }  \; + \;    i n_e \omega }  \; .
\label{sqrt}
   \end{eqnarray} 
In the semiclassical domain $g_4^2 (v) \ll 1$, we use (\ref{bpsmass2}) to obtain:
    \begin{eqnarray}
&&({\rm Dyon \;sum \;at \;large-}L)  \sim e^{i \sigma} e^{ -    \frac{4 \pi v L }{g_4^2}  } \; \;  \sum_{n_e\in \Z} e^{  -  \frac{1}{2}   \frac{v L g_4^2  } {4 \pi} n_e^2   \; + \;    i n_e \omega }  \;  
\label{LLS}
   \end{eqnarray}
 The sum over electric charges in (\ref{LLS}) converges rather fast  for   $\frac{v L g_4^2  } {4 \pi}  \gg1$, i.e., at large-$L$, and the first few terms in the sum are sufficient to produce accurate semi-classical results.   Conversely, this sum converges very slowly if    $\frac{v L g_4^2  } {4 \pi}  \ll 1$, where a more convergent description, as will be described in the next subsection,  emerges---this time in terms of 3d monopole-instantons and twisted monopole-instantons.

 \subsection{3d monopole-instantons at  small $\mathbf{\S^1 \times \R^3}$} 
\label{sec:bos}
 
\begin{figure}[h]
 \begin{center}
\includegraphics[angle=-90, width=5.0in]{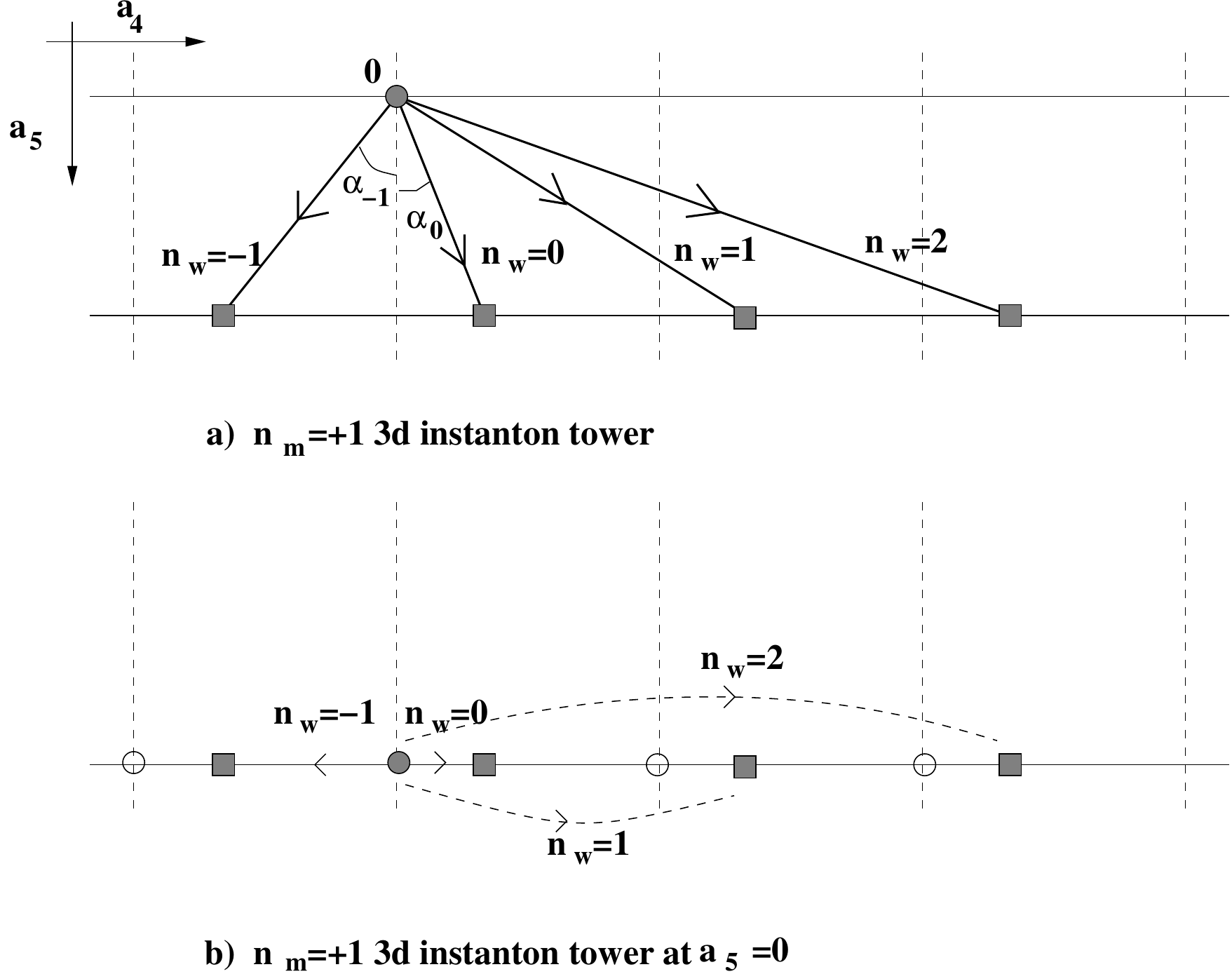}
\caption{Monopole-instanton solutions  for $n_m = +1$ 
are represented by a line with an arrow pointing from the vev of $a_{4,5}$ at the center of the monopole, denoted by a circle, to the vev at infinity, denoted by a square (the vev at the center vanishes and is,  on both pictures, taken to be the origin of coordinates in the $a_{4}/a_5$-plane).  
 Since  $a_4$ is an angular variable on $\R^3 \times \S^1$, a tower of instanton-monopoles of ``winding numbers" labelled by $n_w$ exists. The length of the arrow equals the distance between the vevs at the center of the monopole and infinity, $\sqrt{a_5^2 + a_4(n_w)^2}$, and is proportional to the action of  the corresponding topological defect.  The $n_m = +1$ tower,  shown in the upper figure,  is composed of deformations 
(obtained by turning on $a_5$) of BPS monopole-instantons and $\overline {\rm KK}$ twisted anti-instantons, shown in the lower figure (BPS monopoles have their arrows pointing to the right and $\overline{\rm KK}$-monopoles to the left). The $n_m = -1$ tower,  not shown above, obtained by reverting all the arrows,  is composed of deformations of $\overline{\rm BPS}$ anti-monopole-instantons and KK twisted monopole-instantons. 
 }
  \label{fig:monins}
 \end{center}
 \end{figure}

 Consider again the Euclidean action (\ref{action2}) of the $\N=2$ gauge theory on $\R^3\times \S^1$ and use $A_5,A_6$ to denote the scalar $\phi$, see (\ref{notation3}).  $A_5$ and $A_6$ can be rotated to each other, by using  the symmetries (in what follows we will use this to set $\langle A_6\rangle =  0$), but not to $A_4$ due to the lack of any symmetry relating them. Thus,  we take  only $A_4$ and $A_5$ to have nonzero vevs:
 \begin{equation}
 \label{3dvevs}
 \langle A_5  \rangle = a_5 T_3  \equiv v T_3  = \left[ \begin{array}{cc} 
 \frac{ v}{2} &0 \\
0 & - \frac{ v}{2} 
\end{array} \right],
 \qquad \langle A_4 \rangle=    a_4 T_3  \equiv \frac{\omega}{L} T_3  = \left[ \begin{array}{cc} 
 \frac{ \omega}{2L} &0 \\
0 & - \frac{ \omega}{2L} 
\end{array} \right]
 ~.
     \end{equation}
      Further,   we note that in the compactified theory the vev $a_4$ is actually an angular variable,\footnote{One way one can think of this is that $a_4$ always enters as $\partial_4 + i a_4$ in the Lagrangian and thus can be shifted by $2 \pi\over L$ by relabeling the Kaluza-Klein modes on the circle (or, equivalently, by a ``large" gauge transformation).} $a_4 \equiv a_4 + {2 \pi n \over L}$.
      The $a_4$-$a_5$ slice of the moduli space of the compactified theory is depicted in Fig.~\ref{fig:monins}, where we include all  ``images" along $a_4$ of the chosen vev.

Since $SU(2) \rightarrow U(1)$ by the expectation values (\ref{3dvevs}), the theory has $x_4$-independent finite-action Euclidean monopole-instanton solutions, which are simply the dimensional reduction of the 4d static 't Hooft-Polyakov monopole. 
Consider now the nonvanishing part of the action associated to such a 3d monopole-instanton embedded in $\S^1 \times \R^3$. Take the    monopole solution to be $x_4$ independent and to have  $A_6=0$,  and   $[ A_4, A_5] =0$.
 By using 
 steps similar to   the Bogomolnyi's bound  applied to dyons,\footnote{Earlier, when deriving the dyon equations (\ref{delta}), in the energy functional $E=  \frac{1}{g_4^2} \int_{\R^3}  \tr \left [\vec B^2 + \vec E^2 +    (\vec D A_5)^2  \right] $,  we split the $(\vec DA_5)^2$ term  such that   it compensates both the electric and magnetic field.    This yields the first order dyon equations given in (\ref{delta}); see Ref.~\cite{Weinberg:2006rq} for a review. 
  On $\R^3 \times \S^1$, since $A_4$ and $A_5$ are on different footing (compact vs. noncompact) and cannot be rotated to each other, the $\vec{B}^2$ term in the action should now split  to compensate the two types of scalar  terms, as in (\ref{3dbpsbound}), when applying  Bogomolnyi's technique to our problem.} we obtain, keeping   only the  nonvanishing terms in the bosonic action (\ref{action2}):
    \begin{eqnarray}
    \label{3dbpsbound}
 S=&& \frac{1}{g_4^2} \int_{\R^3 \times \S^1}  \tr \left [\vec B^2 + (\vec D A_4)^2 +    (\vec D A_5)^2  \right] \cr 
 =&&
   \frac{L}{g_4^2}  \int_{\R^3}  \tr \left [ (\vec D A_4 - \sin \alpha  \vec B)^2   +  2  \sin \alpha 
   \vec D A_4   \vec B   + 
      (\vec D A_5 -
   \cos \alpha  \vec B)^2  + 2  \cos \alpha 
   \vec D A_5    \vec B 
   \right] \qquad  \cr
  \geq  &&    \frac{L}{g_4^2}  \int_{\R^3}  \tr   \left[   2  \sin \alpha 
   \vec D A_4   \vec B       +  2  \cos \alpha 
   \vec D A_5    \vec B 
    \right] \qquad  \cr
  =   &&  \frac{L}{g_4^2}  \int_{\R^3}  \vec\partial \;  \tr   \left[   2  \sin \alpha 
    A_4     \vec B       +  2  \cos \alpha 
     A_5     \vec B 
     \right] \qquad   \cr
=&&    \frac{L}{g_4^2}  \left[     \sin \alpha 
    a_4      +   \cos \alpha 
     a_5   \right]     \int_{S^2_{\infty} }     d \vec \Sigma \cdot
      \vec B^3    \cr
 =&&    \frac{L }{g_4^2}  \sqrt{(a_5^2 + a_4^2)}     (4 \pi) \; .
 \end{eqnarray}
In the last step, we assumed that the magnetic charge is $+1$. 
Furthermore, the last equality in (\ref{3dbpsbound}) only holds  when the r.h.s. is minimized with respect to $\alpha$  for the given value of the magnetic charge, i.e. $\alpha$ is given by the vevs  (\ref{3dvevs}) for the $A_5$  and $A_4$ fields as follows: 
\begin{eqnarray}
\label{alpha}
 e^{ i \alpha }=  \frac{ a_5 + i a_4}{\sqrt {a_5^2 + a_4^2}}~.
 \end{eqnarray}
This value of $\alpha$ is denoted by $\alpha_0$ in Fig.~\ref{fig:monins}a. 
The action of the monopole is equal to the minimal value of the r.h.s.  when the solution is BPS saturated:
  \begin{eqnarray}
 && \vec D A_4 - \sin \alpha  \vec B = 0, \nonumber \\
&&      \vec D A_5 -
   \cos \alpha  \vec B=0 .
   \label{second}
 \end{eqnarray}
Few comments are now in order: \begin{enumerate}
\item For $\alpha=\pi/2$, the first equation in (\ref{second})  is the 
 dimensional reduction of the usual instanton equation, $F = \tilde F$ (remembering that $ D_i A_4 = F_{i 4}$, etc.), now obeyed by a self-dual BPS-monopole instanton with $n_m=1$.
 \item For  $\alpha=0$, the second  equation in (\ref{second})  is the  usual Bogomolnyi equation (see~(\ref{delta} with $\delta_{n_e} = 0$) for a magnetic monopole particle with $n_m=1$ in macroscopically four dimensional space-time.
\item For $\alpha=-\pi/2$, the first equation in (\ref{second})  is the 
 dimensional reduction of the anti-instanton equation, $F = -\tilde F$, which is obeyed by an anti-selfdual 
  $\overline {\rm KK}$-monopole-instanton with $n_m = 1$ (recall that the KK-monopole instanton is self-dual and has $n_m = -1$, while the $\overline {\rm BPS}$-monopole-instanton is anti-selfdual and has $n_m = -1$). \end{enumerate}

So far, we have  only addressed the 3d BPS instanton embedded in $\R^3 \times \S^1$ without winding number.\footnote{The existence of the winding (also called ``twisted", or ``KK") monopole-instantons is only possible because  the ``Higgs field" $\sim e^{i L A_4}$ is  compact.
We note that the existence of extra monopole solutions in theories with compact Higgs fields has been noted, but not pursued,  earlier, in the context of maximal abelian projection  \cite{Kronfeld:1987vd}. The advent of $D$-branes  greatly helped  the study of the twisted monopole-instantons, as they appear rather naturally in  string theory brane constructions  \cite{Lee:1997vp}, see also \cite{Kraan:1998pm}.}   
We can generalize the above argument by incorporating the winding number $n_w \in \Z$. Recall that $a_4$ in (\ref{3dvevs}) is really an angular variable on $\R^3 \times \S^1$ and that $a_4 \equiv a_4 \; {\rm mod} \; {2 \pi \over L}$. Monopole-instanton solutions of higher action (``winding number") can be constructed allowing for larger separation between the scalar field eigenvalues at the center of the monopole and infinity---increasing in steps of  $2 \pi n_w \over L$ in the $a_4$ direction, as illustrated on Fig.~\ref{fig:monins}.
Then, repeating the steps in (\ref{3dbpsbound}), we have for winding solutions saturating the BPS bound:
 \begin{eqnarray}
 S_{n_w}=&&    \frac{L}{g_4^2}  \left[     \sin \alpha_{n_w} 
   ( a_4+ \frac{2 \pi}{L}n_w   )      +   \cos \alpha_{n_w}  
     a_5   \right]     \int_{S^2_{\infty} }     d \vec \Sigma \cdot
      \vec B^3    \cr
 =&&    \frac{L }{g_4^2}  \sqrt{a_5^2 + \left( a_4 + \frac{2 \pi}{L}n_w  \right)^2}     (4 \pi),   \qquad 
  n_w \in \Z~,
  \label{small-L-action}
 \end{eqnarray}
 where:
 \begin{eqnarray}
 e^{i \alpha_{n_w}} =  \frac{ a_5 + i  \left( a_4 + \frac{2 \pi}{L}n_w  \right) }{\sqrt {a_5^2 +  \left( a_4 + \frac{2 \pi}{L}n_w  \right)^2 }}~.
 \end{eqnarray}
 The BPS bound on the action (\ref{small-L-action}) is achieved by solutions obeying:\footnote{To avoid a possible confusion  about the $x_4$ independence of the winding solutions, we note that all  winding solutions are $x_4$ independent in a gauge where they asymptote to vacua with $a_4$ vevs differing by $2 \pi n_w/L$, while in a gauge where all solutions asymptote to a vacuum with fixed $a_4$, the winding solutions acquire $x_4$ dependence. The BPS bound is most simply derived in the first gauge.} 
  \begin{eqnarray}
 && \vec D A_4 - \sin \alpha_{n_w} \vec B = 0,\nonumber \\
&&      \vec D A_5 -
   \cos \alpha_{n_w}  \vec B=0 ~.
   \label{alphan}
 \end{eqnarray}
All the instantons  in the   tower have the same magnetic charge, but their topological charges differ by one unit.  Interestingly, this tower is a deformation\footnote{Similar solutions in nonsupersymmetric Yang-Mills-adjoint-Higgs theories are considered in  \cite{Nishimura:2010xa}.} of  the $a_5=0$ theory to non-zero $a_5$ and the instantons pertinent to  this set-up arise from the 
 self-dual  3d-instanton (BPS)  for $n_w\geq 0$  
 and    twisted  ($\overline {\rm KK}$) anti-instanton  for  $n_w\leq -1$.  
The existence of these two types of topological excitations, 3d instantons and the twisted-instantons, is   pertinent to the locally four-dimensional nature of the theory.

 Since the action of a monopole-instanton that saturates the BPS bound (\ref{alphan}) is determined by the vev at infinity, on Fig.~(\ref{fig:monins}), it is geometrically represented by the length of the line\footnote{In the brane description of $\N=4$ SYM these are related to the worldlines of  Euclidean D0-branes \cite{Dorey:2000dt, Dorey:2000qc}.}   stretching between the scalar eigenvalues at the origin and infinity. 
In (\ref{small-L-action}), if we take $a_5=0$ and restrict attention to the 3d instanton for which $n_w=0$, and take $a_4= \pi/L$, which corresponds to a center symmetric background, we obtain from (\ref{small-L-action}), 
 that the action of a 3d instanton in a center-symmetric vacuum is $S_{n_w=0}= \frac{4 \pi^2}{g_4^2}$, which is half that of the 4d instanton.

The 3d monopole-instantons of magnetic charge $+1$ induce amplitudes proportional to  $e^{ - S_{n_w}}  e^{i  \sigma} \times  ({\rm fermion\;  
zero \; modes}) $ in the long distance effective Lagrangian, where $\sigma$ is the dual photon 
defined in (\ref{dual}).  As explained above, there is an infinite tower of such instanton amplitudes contributing to the effective Lagrangian even for an 
instanton with  a given magnetic charge. 
 
\begin{figure}[h]
 \begin{center}
\includegraphics[angle=-90, width=2.0in]{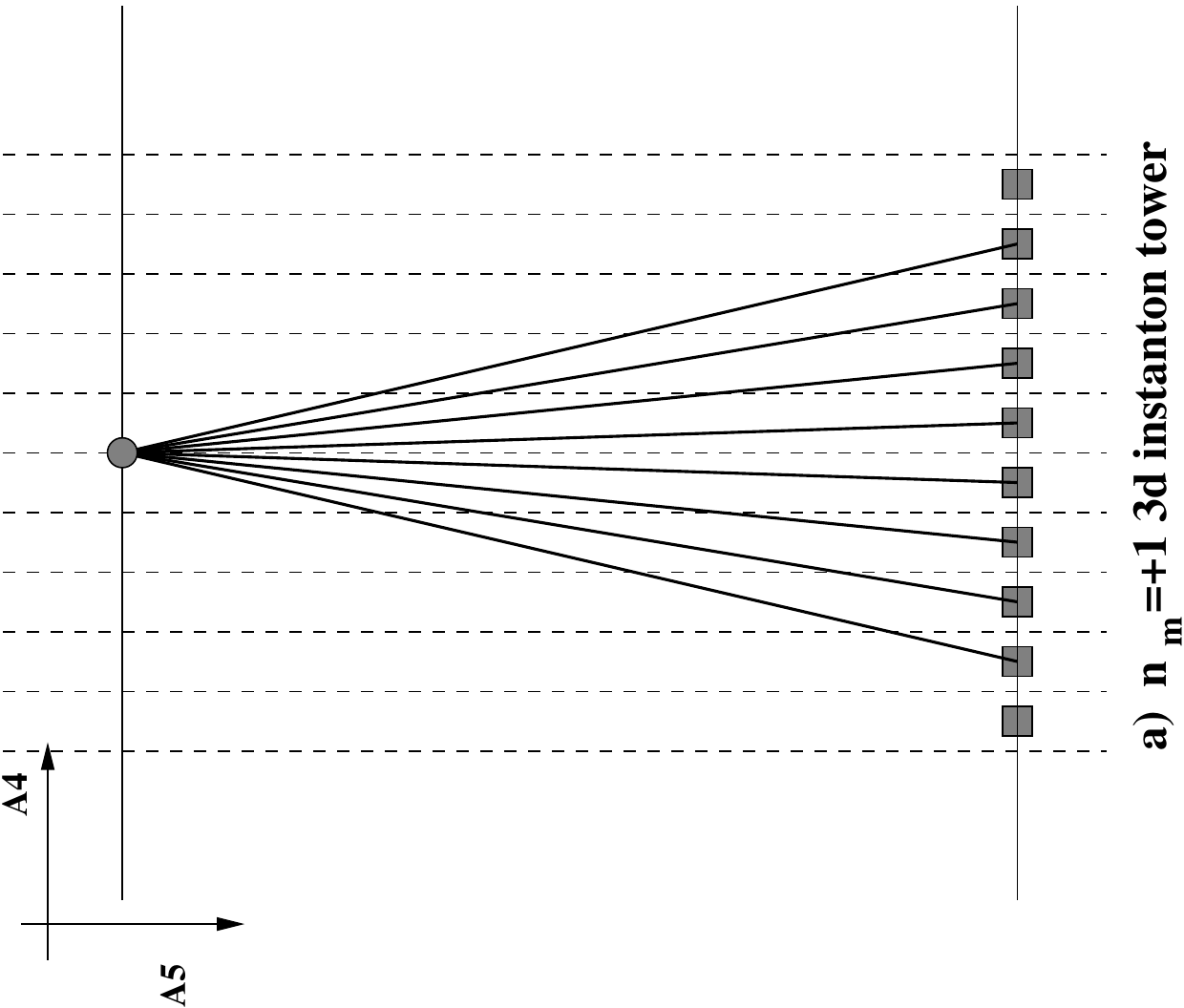}
\caption{3d-instantons in the magnetic charge +1 tower in the regime $a_4 \ll  a_5 \ll a_4/g_4^2$. 
 The tower is composed of deformation of BPS monopole instantons and $\overline {\rm KK}$ twisted anti-instantons. The properties of fermionic zero modes is dictated by the leading $a_5$ dependence for the low winding number  instantons. 
 }
  \label {fig:monins2}
 \end{center}
 \end{figure}

Incorporating the shift due to the $a_5$ vev and winding in the action (\ref{small-L-action}),  and recalling from (\ref{notation3}) that $a_5 = v, La_4 = \omega$, we thus find that the sum of magnetic charge   $+1$  instanton amplitudes   in the small-$L$ domain is, schematically:
   \begin{eqnarray}
&& ({\rm Instanton \;sum \;at \;small-}L )  \sim    e^{i \sigma} F(\omega)  \cr  \cr
&& F(\omega)  \equiv \sum_{{n_w}
  \in \Z} e^{ - S_{n_w}}    =    \sum_{{n_w}
  \in \Z} e^{ -    \frac{4 \pi}{g_4^2}  \sqrt { (vL)^2 +  (\omega + 2 \pi {n_w})^2 }}~,\qquad \qquad 
  \label{sqrt2}
   \end{eqnarray} 
 where  $ F(\omega) $ is the  the sum of the fugacities (i.e., of the $e^{ - S}$ prefactors of the instanton amplitude) and $e^{i\sigma}$ incorporates the long-distance Coulomb interaction between the monopole-instantons.
   In the  regime where $a_5 / a_4 \gg 1$ or $ (Lv \gg 1)$,  the action can be approximated by:
   \begin{eqnarray}
   S_{n_w} \approx \frac{4 \pi v L }{g_4^2}  +   \frac{1}{2} \frac{4 \pi }{Lv g_4^2}  (\omega + 2 \pi {n_w})^2~,
   \label{approx3}
   \end{eqnarray} 
leading to the  asymptotic expression for the small-$L$ instanton sum:
     \begin{eqnarray}
    ({\rm Instanton \;sum \;at \;small-}L) \sim  e^{i \sigma} e^{ -    \frac{4 \pi v L }{g_4^2}  }   \;\;  \sum_{ {n_w} \in  \Z} 
   e^{ -  \frac{1}{2} \frac{4 \pi }{Lv g_4^2}  (\omega + 2 \pi {n_w})^2}~.
   \label{SLV}
   \end{eqnarray} 
  This small-$L$ instanton sum  converges  fast  for  
 $\frac{Lv g_4^2}{4\pi}  \ll 1$, which implies  $\frac{a_5 g_4^2} { a_4}  \ll1$. The combination of this with the  condition ${a_5 \over  a_4 }\gg 1$, used to obtain (\ref{approx3}),  gives:
     \begin{eqnarray}
  a_4 \ll  a_5  \ll \frac{a_4}{g_4^2}   ~,
    \label{hierarchy}
     \end{eqnarray} 
   which can be accomplished at weak coupling. The monopole-instantons that contribute to the sum (\ref{SLV}) are shown on Fig.~\ref{fig:monins2}. 
   
  In the next Section, we will show that the small-$L$ and large-$L$ instanton sums (\ref{SLV}) and (\ref{LLS}) are, in fact, equivalent.

\subsection{3d-instanton/4d-dyon tower Poisson  duality: First pass}
\label{poissondualitysection}

 The statement of Poisson duality, which has been studied in $\N=4$ gauge theories in\cite{Dorey:2000dt, Dorey:2000qc} and in the $\N=2$ context in \cite{Chen:2010yr}, is as follows: the small-$L$  (\ref{LLS}) and large-$L$  (\ref{SLV}) instanton sums,  which, 
  at first sight,  look completely different, are  in fact  {\it equivalent} expressions, and one is the Poisson 
  resummation\footnote{See Section~\ref{poissongeneral}, where we  prove a more general relation, which implies (\ref{Poissond}).}  of the other:
     \begin{eqnarray}
&&   e^{ -    \frac{4 \pi v L }{g_4^2}  +  i \sigma}    \;\;  \sum_{{n_w}  \in   \;  \Z }
   e^{ -  \frac{1}{2} \frac{4 \pi }{Lv g_4^2}  (\omega + 2 \pi {n_w})^2}  \; \times  (\rm four-fermion \; operator)   \cr 
   && =    
       e^{ -    \frac{4 \pi v L }{g_4^2}   +  i\sigma } \; \;  \sum_{n_e \in  \Z }  \sqrt \frac{Lv g_4^2}{8 \pi^2}  \;   e^{  -  \frac{1}{2}   \frac{v L g_4^2  } {4 \pi} n_e^2   \; + \;    i n_e \omega }    \;\times  ( \rm four-fermion \; operator)       
        \label{Poissond}
   \end{eqnarray}
   Both of these are viewed as instanton sums: the first sum, (\ref{SLV}),  over the winding number incorporates 3d monopole-instantons and twisted (or winding) monopole-instantons and their Kaluza-Klein tower.  The second sum, (\ref{LLS}), which should perhaps be called a  dyon sum,  as its salient features are dictated by 4d dyons, is a sum over the electric charges of the dyon tower. In this case, 
   the instantons at large-$L$ are   realized through dyon particles whose worldlines wrap around the large circle.   As discussed earlier, 
the  large-$L$ series converges fast for $Lv g_4^2 \gg 1$ and the small-$L$ series converges fast  for $Lv g_4^2 \ll 1$.

The origin of the Poisson duality (\ref{Poissond}) can be simply understood as follows. In the regime when the expansions  (\ref{bpsmass2})   and (\ref{SLV}) make sense, i.e., at weak coupling and when $Lv\gg 1$, one can interpret the first and second sums in (\ref{Poissond}) in terms of the semiclassical expansion around a classical $x_4$ independent  BPS monopole-instanton solution of zero winding/zero electric charge, respectively, of action $4 \pi v L\over g_4^2$.
 The 4d static monopole solution has a zero mode  associated with global $U(1)_{e}$ transformations, as already mentioned after (\ref{bpsmass2}). Let the corresponding collective coordinate be $\phi\in [0, 2 \pi]$, a compact variable. It is useful to first recall that when quantizing static monopoles as particles on $\R^{3,1}$, $\phi$ is taken to be a function of time (by the coupling of internal and charge symmetries, corresponding to a time-dependent rotation of the monopole), and  has a Lagrangian:
\begin{equation}
{\cal{L}}_\phi =  {I \over 2} \; \dot{\phi}^2~,
\label{phi1}
\end{equation}
where $I$ is the ``moment of inertia" of the monopole of mass $M$ and ``size" $R$, $I =  M ``R^2" ={4 \pi v  \over g_4^2}  {1\over v^2} = {4 \pi \over g_4^2 v}$. This is the Lagrangian of a particle on a circle of unit radius. The corresponding Hamiltonian is:
\begin{equation}
H_\phi = {1 \over 2 I } \; p_\phi^2~,
\label{phi2}
\end{equation}
where the eigenvalues of angular momentum $p_\phi$ are $n_e = 0, \pm 1, \pm 2,...,$ the electric charges of the dyons (recall that the radius of the circle is unity). 

When the path integral on $\R^3 \times \S^1$ is considered,  the ``static" $x_4$-independent solutions become 3d  instanton-monopoles, and after changing variables to collective coordinates in the path integral, we have to integrate over all  functions $\phi(x_4)$, corresponding to the classical paths the particle on a circle can take in a periodic time direction. The net result is that the path integral over the collective coordinate $\phi$ is equal to the partition function of the particle on a circle with Lagrangian ${\cal{L}}_\phi$ (\ref{phi1}) at temperature $1\over L$.

 It is well known that there are two equivalent ways to compute the  partition function  $Z(L)= \tr e^{- L H_\phi}$: 
  One can either compute it in the operator formalism as a sum over states in the Hilbert space of $H_\phi$,  or use the  Feynman  path integral to represent $Z(L)$ as a sum over classical paths of all possible winding numbers.  Let us ignore the  $A_4$ background momentarily. 
If one computes the partition function using the sum over eigenvalues of (\ref{phi2}), one obtains from $\tr e^{- L H_\phi}$ the sum over the electric charges of the dyon tower, the second line in (\ref{Poissond}).
On the other hand, if the partition function is represented as a sum over winding classical paths labelled by winding number $n_w$, $\phi_{n_w}(x_4) = {2 \pi n_w\over L} x_4 + \phi_0(x_4)$, where $\phi_0$ is periodic, one ends up with the small-$L$ representation from the first line in (\ref{Poissond}). In other words,  the two representations of $Z(L)$ are:
\begin{eqnarray}
Z(L) =\tr e^{- L H_\phi} && \equiv  \sum_{n_e \in {\rm Spec}(H_\phi)}   \langle n_e| e^{ - L H_\phi} | n_e\rangle =
 \sum_{n_e \in {\rm Spec}(H_\phi)}  e^{ - \frac{L n_e^2}{2I}}  \cr \cr
&&  \equiv   \int\limits_{\phi(0)= \phi(L)({\rm mod} \;2 \pi) } D\phi(\tau) e^{- \int_0^{L} d\tau   {I \over 2} \; \dot{\phi}^2} \cr \cr
&& = 
 \sum_{n_w \in \pi_1(S^1) } \sqrt\frac{2 \pi I}{L} e^{ - \frac{I}{2L}(2 \pi n_w)^2},  
 \label{ZL}
\end{eqnarray}
where the prefactor on the second line in (\ref{Poissond}) is related to the properly normalized path integral over the periodic $\phi_0$. Since ${1\over I} = {g_4^2 v \over 4 \pi}$, we immediately see the equivalence of  (\ref{ZL})  to (\ref{Poissond}).

The coupling to $A_4 = \omega/L$ can be similarly understood by considering the static monopole in an external $A_0$ field, which modifies (\ref{phi1}, \ref{phi2}), to ${\cal L}_\phi = {I \over 2} (\dot{\phi} + i A_0)^2$ and $H = {1 \over 2 I} p_\phi^2 - i A_0 p_\phi$, respectively, and a subsequent continuation to Euclidean space. From the thermal analogy, it is also clear that at low temperature  (large-$L$), the partition function converges faster if given as a sum over the quantized eigenvalues of $H$, while at high temperature (small-$L$) it is better represented as a sum over classical paths.

Going back to  the monopole-instanton sums (\ref{Poissond}),  the Poisson duality means  that we can describe the system most conveniently in terms of the few leading topological defects in the respective sums wherever they converge fast. However, this is 
an issue of convenience and the physical content of the two sums is identical. Nevertheless, we would have to do much work to describe the  small-$L$  phenomena by using the large-$L$ degrees of freedom and vice versa.
The Poisson resummation also instructs us that the electric charge and winding number  are 
 dual variables. As is clear from the above discussion, in this system, the  quantization of the dyon electric charge may also be thought as the dual of the quantization of  winding number of 3d monopole instantons.\footnote{In the string theory embeddings this follows from momentum/winding number $T$-duality along $\S^1$ \cite{Dorey:2000dt, Dorey:2000qc}.}

In the presence of massless fermions, the instanton amplitudes are associated with a number of fermionic zero modes.  In general, the fermionic  zero modes rotate  as a function of the two adjoint Higgs vevs,  $a_4$ and $a_5$, and vary  throughout the    moduli space.  We postpone the detailed  discussion of the preserved supersymmetries and the fermion zero mode structure  to Appendix~\ref{chiralityappendix} for conciseness.\footnote{However, this  is an integral part of the discussion. The reader who is  not familiar with this line of reasoning should study the derivations in Appendix~\ref{chiralityappendix}.} For now, we only note that  in the  limit that we derived the  Poisson duality, $\alpha_{n_w} \approx 0$  for a large number of instantons in the 3d tower (see Fig.~\ref{fig:monins2}); similarly  in the semi-classical and non-relativistic limit of dyons, giving rise to the dense band of states  shown in  Fig.~\ref{fig:dt}, we have  $\delta_{n_e} \approx 0$. In these two regimes,  the wave functions of the fermion zero modes of the two towers are identical to leading order, as follows from  eqns.~(\ref{ins-towerP}) and (\ref{dyon-towerP}). This demonstrates  that the Poisson duality (\ref{Poissond}) is not harmed by the inclusion of fermionic zero modes.

 \subsection{General Poisson duality, or  4d dyon spectrum from 3d instanton sums}
  \label{poissongeneral}
 
 In this subsection, we demonstrate that    the  relativistic  version of the  BPS spectrum (\ref{bpsmass1}) and the central charge  formula  $Z_{(n_m, n_e)}$ for the magnetic dyon tower in 4d
 can be extracted from  the sum over the tower of 3d instantons.   This procedure  is parallel to the well-known textbook example on the extraction of the energy spectrum of $O(N)$ rigid rotator from the path integral  formulation of quantum mechanics.  See,  for example, Ref.~\cite{ZinnJustin:2002ru}.    
 We present the calculation in detail for pedagogical reasons.

Consider first the following small-$L$ sum over the instanton tower  (\ref{sqrt2}) labeled by the winding number $n_w$ and magnetic charge $n_m=1$. The winding number $n_w \in \Z$ is the integer part of the topological charge of the corresponding  monopole-instanton in the tower. The sum (\ref{sqrt2}) is a  generalized fugacity $ F(\omega) $: 
      \begin{eqnarray}
      \label{fugacity}
 F(\omega) =  \sum_{n_w
  \in \Z} e^{ - S_{n_w}}    =      \sum_{n_w
  \in \Z} e^{ -    \frac{4 \pi}{g_4^2}  \sqrt { (vL)^2 +  (\omega + 2 \pi n_w)^2 }} ~,
  \end{eqnarray}
This is  clearly a periodic function of holonomy $\omega$, 
   $ F(\omega + 2 \pi)=  F(\omega )$. 
   Introduce its Fourier transform:
    \begin{eqnarray}
    \label{Fourier1}
F(\omega ) = \sum_{n_e \in \Z}  F_{n_e } e^{i \omega n_e }~,
 \end{eqnarray} 
  where  $e^{i \omega n_e }$ is the canonical coupling of the background gauge field $A_4$ to electric charge $n_e$, $e^{i \int_{\S^1} n_e A_4} \equiv e^{i \omega n_e }$. Remembering that the fugacity we are calculating appears in the combination $e^{i \sigma} F(\omega)$,   we deduce that the  Fourier coefficients $F_{n_e }$ are associated with dyons of magnetic and electric charges   $(1, n_e)$.  $F_{n_e }$ should have an interpretation as the Boltzman weight, just like it was the case in our non-relativistic discussion in 
Section~\ref{poissondualitysection},  but now for a relativistic particle  with  
Hamiltonian $H_\phi = \sqrt { M_{(1,0)}^2 + v^2 p_\phi^2}$ and  eigenspectrum $M_{(1,n_e)}=  v
  \sqrt { \left( \frac{4\pi}{g_4^2}\right)^2  +  n_e^2}$.    
We will indeed see that   $F_{n_e } \sim e^{-LM_{(1,n_e)}}$.  To this end, let us invert   (\ref{Fourier1}): 
     \begin{eqnarray}
     \label{intF}
 F_{n_e }   &=&  \int_{0}^{2 \pi}  
\frac{d \omega}{2 \pi}   F(\omega )  e^{-i \omega n_e } \nonumber \\&=&  \sum_{n_w 
  \in \Z}  \int_{0}^{2 \pi} \frac{d \omega}{2 \pi} e^{ -    \frac{4 \pi}{g_4^2}  \sqrt { (vL)^2 +  (\omega + 2 \pi n_w)^2 }} 
 e^{- i \omega n_e} \\
&=&    \int_{-\infty}^{\infty}  
\frac{d \omega}{2 \pi}   e^{ -    \frac{4 \pi}{g_4^2}  \sqrt { (vL)^2 +  \omega^2 }}  e^{-i \omega n_e }~.  \nonumber
 \end{eqnarray} 
We will see below that   $F_{n_e}$ equals the contribution of a single dyon pseudoparticle of electric charge $n_e$. Thus, the middle line of (\ref{intF})  expresses the single dyon contribution as    a linear combination  of contributions of an infinite number of winding monopole-instantons.
 
 We now perform the integral in (\ref{intF}) 
by using the change of variables
$
  \omega= vL \sinh t$, $d\omega= vL \cosh t  dt$:
    \begin{eqnarray}
      F_{n_e } 
  = \frac{vL}{4 \pi}   \int_{-\infty}^{\infty}    e^{ - \frac{vL}{2}\left( (\frac{4 \pi}{g_4^2}  + in_e )  e^t + 
    (\frac{4 \pi}{g_4^2}  - in_e )  e^{-t}  \right)}  (e^t + e^{-t}) dt~,
 \end{eqnarray} 
and then split the integrand in two. In the first one, we use  $e^t=u$ and in the second---$e^{-t}=u$.
Then, we have $F_{n_e} = \frac{vL}{4 \pi}   (F_{1n_e } + F_{2n_e })$, where    $F_{2n_e }= F_{1n_e }^*$, and obtain:
    \begin{eqnarray}
      F_{1n_e } 
   && =   \int_{0}^{\infty}    e^{ -\frac{vL}{2}\left( (\frac{4 \pi}{g_4^2}  + in_e )  u + 
    (\frac{4 \pi}{g_4^2}  - in_e ) \frac{1}{u}  \right)}  du   = 2 \sqrt \frac{  (\frac{4 \pi}{g_4^2}  - in_e )} {  (\frac{4 \pi}{g_4^2}  + in_e )}  \; 
     K_1 \left( vL  \sqrt {  \left( \frac{4 \pi}{g_4^2} \right)^2 +  (n_e )^2 } \right) \cr 
     && =  2 e^{-i \delta_{n_e} }   K_1 (L M(1, n_e )),
 \end{eqnarray} 
where $K_1$ is the Bessel function, $M(1, n_e )$ is the mass of the  dyon with electric charge $n_e $, and ${\delta_{n_e} }$ is the angle defined in finding the dyon equations of motions (\ref{delta}).    Thus,  $F_{n_e }$ becomes:
  \begin{eqnarray}
      F_{n_e } =   \frac{vL}{4 \pi} \left(    2 e^{-i \delta_{n_e }}   +  2 e^{i \delta_{n_e} }  \right)  K_1 (L M(1, n_e ))  
     =  \frac{vL}{ \pi}  \cos\delta_{n_e } \; \;   K_1 (S(1, n_e )),
     \label{Fne1}
 \end{eqnarray} 
 where $S(1, n_e )$ is the action  of the  dyon (\ref{dyonaction}). 
 
 Substituting (\ref{Fne1}) back into (\ref{Fourier1}), the generalized Poisson duality is: 
 
 \newpage
     \begin{eqnarray}
       \label{generalizedPoisson1} 
    \sum_{n_w
  \in \Z} e^{ -    \frac{4 \pi}{g_4^2}  \sqrt { (vL)^2 +  (\omega + 2 \pi n_w)^2 }}  =&&   \sum_{n_e \in \Z}    \frac{vL}{ \pi}  \cos\delta_{n_e} \; \;   K_1 (LM(1, n_e))    e^{i \omega n_e}  \cr
  =&& \sum_{n_e \in \Z}   \frac{vL}{ \pi}  \cos\delta_{n_e} \; \; \sqrt \frac{\pi}{2 LM(1, n_e)}     e^{- LM(1, n_e)}     e^{i \omega n_e} ~.
 \end{eqnarray} 
  In the second line, we used the  $x \rightarrow \infty$ asymptotic expansion  $K_1(x) \approx \sqrt \frac{\pi}{2 x} e^{-x}$.  As promised, the sum  over the 3d instanton tower is Poisson dual to the sum over the spectrum of   the 4d dyon particle tower. The 4d spectrum of dyons can easily be obtained through the formula:
  \begin{equation}
  \label{spec5}
M(1, n_e) = - \lim_{L \rightarrow \infty} \left( \frac{1}{L} \log F_{n_e} \right) ~.
 \end{equation}
 This formula  generalizes to dyons with higher magnetic charges as well.   
  We note that in (\ref{spec5}), the l.h.s. is a  4d quantity while the r.h.s.  is given as an infinite sum over 3d contributions comprising $F_{n_e}$, see (\ref{intF}), given in terms of sum over paths.\footnote{   
 This is actually a special and particularly simple case of a more  general Poisson duality formula, formulated  semi-classically by Gutzwiller and  later generalized to an exact relation by Selberg \cite{bookchaos}. 
 The winding number is associated  with the first homotopy group and  the lengths of closed geodesics are  in one-to-one correspondence with the homotopy classes in the compact space. 
 The electric charges provide spectral data about the Hamiltonian, which describes the $U(1)_e$ collective coordinate of the monopole particle. Gutzwiller  trace formula  is a relation between the lengths of closed geodesics  and the quantum spectrum of a Hamiltonian, and applies also to spaces with 
 non-abelian isometry groups.  It may be interesting to examine if there is anything of significance in this relation for the monopole physics  and whether it has connections to a non-abelian generalization of T-duality in string theory.}

   In the semi-classical domain discussed in Section~\ref{poissondualitysection}, we  assumed $\cos\delta_{n_e} \approx 1$ for the low-lying dyon band, and used the non-relativistic approximation for the dyon mass of eqn.~(\ref{bpsmass2}) and the expansion  (\ref{approx3}) for the twisted instanton action.  Inserting these expansions, on both sides, we obtain     
 exactly (\ref{Poissond}) of Section \ref{poissondualitysection}; the prefactors also match  exactly.

 \subsection{Digression: The meaning of the sum}
 \label{digression}
 
 One might ask what quantities in the $\N=2$ theory does the sum in the general
 Poisson duality formula  (\ref{generalizedPoisson1}) relate? In fact, this question was already addressed in \cite{Chen:2010yr}. It turns out that the r.h.s. of (\ref{generalizedPoisson1}) is proportional to the large-$L$ dyon tower contribution to the  K\" ahler potential, where the complex moduli $v = (A_5 + i A_6)/\sqrt{2}$ and $ \sigma - i b = \sigma - i {4 \pi  \omega \over g_4^2}$, recall (\ref{notation3}), are used to parameterize the hyper-K\" ahler manifold. More explicitly, the contributions of the dyon tower to the K\" ahler potential are given by:\footnote{Up to an overall constant, which is the same in (\ref{Kdyon}) and (\ref{Kwinding}).}
 \begin{equation}
 \label{Kdyon}
 K_{dyon} = {1 \over \sqrt{2} \pi^{3\over 2} L^{3 \over 2} |v|^{1\over 2} }  \;  \sum\limits_{n_m = \pm 1} \sum\limits_{n_e  \in \Z} 
 { 
 e^{ - L |v| \sqrt{ \left(4 \pi \over g_4^2\right)^2 + n_e^2} + i \omega n_e + i \sigma n_m } 
 \over  
 \left[\left(4 \pi \over g_4^2\right)^2 + n_e^2\right]^{1 \over 4}
 }~,
 \end{equation}
while the Poisson resummed K\" ahler potential, expressed now as a sum over the contributions of winding solutions, and thus appropriate at small $L$, is:
\begin{equation}
\label{Kwinding}
K_{winding} = K_{dyon} = {1  \over \pi L^2 |v|} \; \sum\limits_{n_m = \pm 1} \sum\limits_{n_w  \in \Z}  
  e^{ - {4 \pi L \over g_4^2} \sqrt{|v|^2 + 
\left( \omega + 2 \pi n_w \over L\right)^2} + i \sigma n_m} ~.
\end{equation}
Note that $K_{winding} = K_{dyon}$ is precisely equivalent our equation (\ref{generalizedPoisson1}) in the $g_4^2 \rightarrow 0$ limit (i.e., with $\cos \delta_{n_e} \approx 1$).

The sum (\ref{Kdyon}) over the dyon tower contributions to the K\" ahler potential was obtained in \cite{Chen:2010yr} by solving the  equations  \cite{Gaiotto:2008cd} obeyed by the hyper-K\" ahler metric  on the moduli space of the $\N=2$ theory on $\R^3 \times \S^1$ iteratively for weak coupling. Thus, the expressions for the K\" ahler potentials (\ref{Kdyon}) and (\ref{Kwinding}) are valid in the limit $|v| \gg \Lambda_{\N=2}$, but for arbitrary values of $|v| L$ . Note that in the regime where (\ref{Kdyon}), (\ref{Kwinding}) are valid,  $|v| \Lambda_{\N=2} \gg 1$, there is no ``wall-crossing" and that this  regime is different from the large-$L$ (but arbitrary coupling) regime $|v|L  \gg 1$ studied in \cite{Gaiotto:2008cd}. We also note that in \cite{Chen:2010yr}, instead of the K\" ahler potential (\ref{Kwinding}) for $v$ and $\sigma - i b$, the  K\" ahler metric component $g_{v v^*}$ was actually given, but it is a simple matter to check their equivalence.

 We see that the Poisson duality explained in Section~\ref{poissondualitysection} is actually more generally valid. However, it is difficult to give a semiclassical test of this more general duality, since the fermion zero modes of the different winding and dyon backgrounds are different (see Appendix \ref{chiralityappendix})  and demonstrating the Poisson duality would entail showing that the sums over the various four-fermion operators agree (Ref.~\cite{Chen:2010yr} only provided a semiclassical test of the duality in the simplifying regime where the four-fermion operators of the two towers are identical, just as we did near eqn.~(\ref{Poissond})).  
However, the derivation starting from the K\" ahler potential  or K\" ahler metric    \cite{Chen:2010yr}, which assumes the validity of the ``wall-crossing" formula \cite{Gaiotto:2008cd}, shows that the duality is clearly valid---the four-fermion terms follow from the metric and the relevant zero modes appearing in the four-fermion amplitude  adjust themselves as the moduli and quantum numbers of the relevant instantons vary.

To summarize, in  Sections~\ref{poissondualitysection}, \ref{poissongeneral}, and \ref{digression},  we argued that  in the semiclassical regime along the path $C$ of Fig.~\ref{fig:ourpath}, the topological defects responsible for confinement  at small and large $L$ are related by Poisson duality. 
At large $L$, the $(n_m, n_e)=$ $(1,0)$ and $(1,1)$ members of the dyon tower and their antiparticles  survive continuation to the strong-coupling $u \sim 1$ domain and become massless at the monopole and dyon points, respectively. At small $L$, it is the 
lowest action winding monopole-instantons, $(n_m, n_w)=$ $(1,0)$ and $(-1,-1)$  and their anti-instantons  that play a crucial role in the dynamics of confinement. Next, we discuss these confinement mechanisms in turn.

\section{Mass deformed Seiberg-Witten theory on $\mathbf{\S^1 \times \R^3}$}
\label{massdeformed}

In the next Section \ref{largeLSW}, we recall the description of the monopole and dyon condensation in the mass-perturbed theory at large $L$, $L \gg \Lambda^{-1}_{\N=2}$. 
The local dynamics at distances smaller than $L$ is  the one of 4d SW-theory. 
 If we are interested in physics at distances larger than the $\S^1$ size $L$,  then the dynamics needs to be  described  by the dimensional reduction of the relevant SQED.  We give a description based on various 3d infrared  dualities.\footnote{Section~\ref{largeLSW}  uses a 4d $\N=1$ (or, equivalently, 3d $\N=2$) formalism for the compactified theory.  The   reasoning entering   the description of the asymptotically  long-distance physics  is based on  symmetries and  known 3d dualities.  A good background material for this part is found in the lecture notes \cite{Strassler:2003qg}. Our description will be brief since it does not have significant overlap with the rest of our discussion.} In  Section \ref{sec:smallLdynamics}, we elucidate the  role of the  monopole-instantons and magnetic bions   in the less well-known small-$S^1$ confinement and chiral symmetry breaking  dynamics. 

\subsection{The large-$\mathbf L$ regime}
\label{largeLSW}

On $\R^4$, Seiberg and Witten (SW) showed that  the theory around any one of the  two singular points in moduli space  can be described as $d=4$,  $\N= 2$ SQED. These points are usually referred to as monopole ($u=1$) and dyon ($u=-1$) points. Near these points, the $(1,0)$ and $(1,1)$ dyons, respectively, are described by hypermultiplets charged under the electromagnetism of the relevant dual photon multiplet, $\cal{A}_D$; we note again that there is no globally valid effective field theory that describes both the monopole and dyon point in moduli space. 
Precisely at the monopole/dyon points, these hypermultiplets become massless.

When a mass perturbation reducing the $\N= 2$ supersymmetry  to $\N= 1$ is added, SW theory exhibits confinement due to the condensation of the monopoles or dyons. 
Explicitly, using $\N=1$ superfields, the superpotential  of SQED near either of the two singular points has the form:
\begin{eqnarray}
W = \sqrt 2 A_D Q \widetilde Q + m u(A_D)
\label{effL}
 \end{eqnarray}
 where $Q$ and $ \widetilde Q$ are chiral monopole multiplets  with charges, $\pm 1$, respectively 
 under the  electromagnetism of  the  dual photon vector multiplet, ${\cal  A}_D$.  
   $A_D$ is the   $\N=1$  scalar component of  ${\cal  A}_D$ and the first term in (\ref{effL}) is required by  
     $\N=2$ supersymmetry of the $m=0$ theory. 
Near the monopole point, 
  \begin{eqnarray}
 A_D \approx c(u-1)~.
  \end{eqnarray}
 The vacuum of the theory can be found by solving simultaneously the $D$- and $F$-term equations. Vanishing of the dual $U(1)$ $D$-term gives
 $ D=   Q^{\dagger}  Q - \widetilde Q^{\dagger}   \widetilde Q=0$.
  The critical point of the superpotential is
  $\frac{\partial W}{\partial u} =  \frac{\partial W}{\partial Q} = 
 \frac {\partial W}{\partial  \widetilde Q}=0 $ and yields: 
 \begin{eqnarray}
  Q \widetilde Q = -\frac{ m}{ c \sqrt 2} , \qquad  c(u-1)\widetilde Q =0, \qquad  c(u-1) Q =0 ~.
 \end{eqnarray}
This means that monopoles, which are massless particles in the  $\N=2$ theory,
become tachyonic once   the   mass perturbation $m \ll \Lambda_{\N=2}$ is added. The mass-perturbed SW theory exhibits a magnetic Higgs mechanism, has a mass gap and two isolated vacua (corresponding to either the monopole or dyon points). 
Confinement of electric charges is due to the Abrikosov-Nielesen-Olesen strings of the dual abelian Higgs model (SQED) describing the long-distance dynamics at the monopole or dyon point.\footnote{See \cite{Douglas:1995nw} for  detailed calculations for an $SU(N)$ gauge group.} Restoring explicit factors of $\Lambda_{\N =2}$, the monopole (dyon) condensate is $\langle Q \rangle ~ \sim \sqrt{m \Lambda_{\N =2}}$ and the mass gap $M^2$ (i.e., the masses of the dual photon vectormultiplet and the monopole hypermultiplet) and  the confining string tension  $\Sigma$ are given  by: 
\begin{equation}
M^2  \sim ~e_D^2  m \Lambda_{\N=2}, ~ ~~\Sigma  \sim m \Lambda_{\N=2},
\label{swstringtension}
\end{equation}
where $e_D$ is the IR-free dual photon coupling:
 \begin{equation}
 \label{4ddualphotoncoupling}
 e_D^2 \sim  { 16 \pi^2 \over  \log{\Lambda_{\N=2}\over m}}~,
\end{equation} 
 which, in the mass-perturbed theory,  is ``frozen" due to $\langle Q \rangle \ne 0$.

As already mentioned in the Introduction, while the effective SQED descriptions of the monopole and dyon points are  valid at $L \Lambda_{\N=2} \gg 1$, we do not know how continue them to the small-$L$ regime. First, recall that the monopole/dyon point SQEDs are effective theories, valid at energies below $\Lambda_{\N=2}$, and upon compactification, once $L \Lambda_{\N=2}$ becomes of order unity, the effective theory description breaks down since the IR free dual photon coupling (\ref{4ddualphotoncoupling}) becomes strong at the scale $1/L$.  Second, as will become clear from the small-$L$ description of the dynamics in  Section \ref{sec:smallLdynamics}, for $0 < L \Lambda_{\N=2} \ll 1$, there is a globally valid weak-coupling description of the two vacua of the mass-perturbed SW theory, while such a global description is lacking at large $L$. Thus, as $L$ is decreased, the two mutually nonlocal descriptions have to merge into a single one and it is beyond our current knowledge to describe this in field theory.

  A more modest task is to consider 
    a compactification of the SW theory on $\R^3 \times \S^1$ of a radius such that $L \gg 1/\Lambda_{\N=2}$.   Then,  the local dynamics is essentially four dimensional and the SQED descriptions   near the monopole or dyon points in moduli space considered above are certainly valid. If we are interested in physics at distances larger than the $\S^1$ size $L$, we can describe the dynamics by the dimensional reduction of the relevant SQED. For this purpose, SW theory near the 
   monopole/dyon points reduces to $\N=4$ SQED in 3d---the dimensional reduction of the 4d dual photon SQED with superpotential (\ref{effL}); for now, we ignore the mass perturbation. 
The 3d gauge coupling at  the UV cutoff of the 3d theory is $e_3^2 = e_D^2(L)/L$, where $e_D^2$ is the 4d dual photon coupling.

A convenient way to study  the $\N=4$  3d SQED is  to start with  $\N=2$  3d SQED.   
By adding a gauge-singlet chiral multiplet  $\Psi$  with  a superpotential 
$W = \sqrt 2 \Psi  Q \widetilde Q$ (and a normalization of the kinetic term appropriate to an $\N = 4$ multiplet), one obtains   $\N=4$ SQED in 3d. 
As we will see shortly, this way of realizing the $\N=4$ SQED in 3d is useful, because it will allow us to explore the IR dynamics through  a chain of known 3d dualities; see \cite{Strassler:2003qg} for a review.  This is also necessary, since 
the dimensional reduction of the 4d IR free SQED with massless charged fields is not IR free and exhibits nontrivial dynamics.

Let us start with the  $\N=2$ 3d SQED as there is a  duality relating it to a global (non-gauge) theory.  
The vacuum of the $\N=2$ 3d SQED theory is one complex dimensional and is parameterized, asymptotically, 
by three chiral superfields  $\{V_{+} , V_{-}, M \}$; the first two
 $V_{+} \sim e^{ \Phi^D} =  e^{ \phi^D +  i \sigma^D }$, 
$V_{-} \sim e^{ -\Phi^D} $
  labeling the Coulomb branch (which splits in two\footnote{Here, $\phi^D$ is the component of the 4d dual photon gauge field along the compact direction and $\sigma^D$ is  the 3d scalar  ``dual photon" of the dimensionally reduced  4d dual photon of SW theory (we do not see a way to avoid this potentially confusing terminology).}), and a meson superfield $M=  Q \widetilde Q$ denoting the Higgs branch.  Ref.\cite{Aharony:1997bx} argues that the three branches meet at the origin of the moduli space and they are in fact related by a triality exchange symmetry  
  $ V_{+} \rightarrow V_{-} \rightarrow M  \rightarrow V_{+}$. 
   Ref.\cite{Aharony:1997bx}  also  provides evidence that   $\N=2$  SQED in 3d is dual to a Wess-Zumino (WZ) model with superpotential:
\begin{equation}
W= V_{+}V_{-} M
\label{3dw1}
\end{equation}
 in the infrared,\footnote{With canonical 3d normalization of the fields, the dimensionful coefficient in (\ref{3dw1}) should be $\sim e_3$ (although this is irrelevant for the long-distance dynamics of the WZ model, which is described by a 3d strongly-coupled ``Wilson-Fisher" SCFT); the same   factor makes up  the dimensions in (\ref{3d1})   as well.}  where $V_{+}, V_{-}$, and $ M$ are unconstrained chiral superfields.
 
Now, we can study $\N=4$ 3d SQED by adding the image of the superpotential 
$W = \sqrt 2 \Psi  Q \widetilde Q$
in the  $V_{+}V_{-} M$  WZ model. Since   $ Q \widetilde Q $ is the meson $M$,  $W = \sqrt 2 \Psi  Q \widetilde Q \equiv   \Psi M$, and thus, 
the superpotential of the  $V_{+}V_{-} M$ theory is deformed to: 
\begin{equation}
W= V_{+}V_{-} M + \Psi M~.
\label{3d1}
\end{equation}
The last term gives mass to $M$ and $\Psi$ and removes them from the effective theory.  This leaves us with $V_{+}$ and $V_{-}$ with no superpotential. The IR-dual description of 3d $\N=4$ SQED is thus a free field theory. $V_{+}$ and $V_{-}$ parametrize a four-real dimensional manifold;  this is  a local patch of a hyper-Kahler manifold with quaternionic dimension one (presumably, the Atiyah-Hitchin manifold \cite{Seiberg:1996nz,Dorey:1997ij}). 

Finally, let us study the effect of the mass deformation in SW theory (adding this perturbation in the long-distance 3d theory assumes that  the mass gap (\ref{swstringtension}) obeys $M \ll e_3^2$, since  the IR-dual   (\ref{3d1}) is only valid at scales well below $e_3^2$; recall also that  $e_3^2 \ll 1/L$).   Based on our understanding of SW theory on $\R^4$,  we expect that the mass-perturbed theory should have a mass gap.  Indeed, the  mass perturbation of  superpotential (\ref{3d1}) is: 
 \begin{equation}
W= V_{+}V_{-} M + \Psi M + \epsilon \Psi,
\end{equation}
with $\epsilon \sim \sqrt{m \Lambda_{\N=2}}$ ,  
 whose  critical points   lead to a unique vacuum at  $M= -\epsilon, V_{\mp}=0,  \Psi=0$. It is easy to see that all 
 excitations acquire mass and the theory is gapped, just as its 4d counterpart. 
 
 These consistency checks  are the most we can achieve using the 3d language, since we are considering   distances $\gg L$ and the 4d language of massless monopoles becoming tachyonic and condensing is not   the appropriate one here.

 \subsection{The small-$\mathbf L$ regime}
\label{sec:smallLdynamics}
 
 We now consider the limit of small circle size.   In the small $\S^1 \times \R^3$ domain of the $\N=2$ theory, 
  there are two types of 3d monopole instantons as well as their Kaluza-Klein towers, which were described in Section \ref{sec:bos}. 
For definiteness, anticipating the applications to the mass-deformed theory, we will  consider a point in moduli space where $a_5 = 0$, $a_4 \ne 0$. Further, assuming  $a_4 \sim \pi/L \gg \Lambda_{\N=2}$, weak-coupling semiclassical methods are clearly applicable. In the region of moduli space chosen, the angle $\alpha$ of eqn.~(\ref{alpha}) equals $\pm \pi/2$ and the monopole instanton solutions are (anti-)self-dual, as explained after eqn.~(\ref{second}). The action of the solutions of magnetic charge $\pm1$ and arbitrary winding is given in (\ref{small-L-action}). 

The self-dual solutions are the BPS monopole-instanton  $(n_m, n_w) = (1,0)$ of action $b =  { 4 \pi \over g_4^2 } L a_4 ={ 4 \pi \over g_4^2 } \omega$ (see (\ref{small-L-action}) with $n_w=0$,  (\ref{notation3}), and note that we take $0 \le \omega \le 2 \pi$)). Below, we will denote the 't Hooft operator generated by the BPS monopole-instantons by ${\cal{M}}_1$.
The other lowest action self-dual solution is the KK monopole and $(n_m, n_w) = (-1,-1)$, which has action action ${8 \pi^2 \over g_4^2} - b = { 4 \pi \over g_4^2 } (2 \pi - \omega)$  (see (\ref{small-L-action}) with $n_w=-1$). The 't Hooft vertex associated with this instanton will be denoted  by ${\cal{M}}_2$. Since we are at small $L$ here, it will be sufficient to consider only the contributions of these two lowest action solutions. These solutions have four fermion zero modes each (and, since they are self-dual, their zero modes are chiral in 4d sense, see also Appendix \ref{chiralityappendix}).  To summarize, the magnetic and topological charges $(Q_m, Q_T)$ associated with these instantons (and the corresponding anti-instantons) are:
\begin{eqnarray}
&&{\cal M}_{1} : (+1,  +\half) 
\qquad 
{\cal M}_{2} :  (-1,   + \half)  \cr  \cr
&&\overline {\cal M}_{1} : (-1,  - \half)   \qquad 
\overline {\cal M}_{2} : (-1,  - \half)~.
\end{eqnarray}
Note that the topological charge equals $1\over 2$ only at the center-symmetric point $\omega = \pi$.

These monopole-instantons generate  four-fermi interactions in the effective long-distance lagrangian, of the form:
  \begin{eqnarray}
  \label{4fermi}
 L_{\rm 4F} = {\cal M}_{1} + {\cal M}_{2}  + \overline {\cal M}_{1} + \overline {\cal M}_{2}~,
   \end{eqnarray} 
where 
the amplitudes associated with these instanton events (and their  anti-instantons) are given by:
\begin{eqnarray}
&&{\cal M}_{1} =e^{ - b +i \sigma} \lambda \lambda \psi \psi  \qquad 
{\cal M}_{2} =  \eta e^{ + b -i \sigma} \lambda \lambda  \psi \psi , \qquad \eta 
\equiv e^{2 \pi i \tau} =   e^{- \frac{8\pi^2}{g_4^2} + i \theta}  \cr  \cr
&&\overline {\cal M}_{1} = e^{ - b - i \sigma} {\bar \lambda}  \bar \lambda  \bar \psi \bar  \psi \qquad 
\overline {\cal M}_{2} =  \bar \eta   e^{ + b + i \sigma} \bar  \lambda \bar \lambda  \bar \psi \bar \psi\; .
\label{2amplitudes}
\end{eqnarray}
  The combination ${\cal M}_{1} {\cal M}_{2}$ with eight zero mode insertions is the 4d-instanton amplitude and has the form  $\eta (\lambda \lambda \psi \psi)^2$     given in  (\ref{4d-instanton}). This amplitude, as noted in Section~\ref{sec:review}, reduces the chiral $U(1)_R$ symmetry down to $\Z_8$, which is a true  anomaly-free symmetry of the quantum theory. Clearly, the chiral four-fermion operators in (\ref{2amplitudes}) flip sign under $\Z_8$. This implies that for the operators 
in (\ref{2amplitudes}) to remain invariant, one needs a discrete shift in the $\sigma$ field, of the form:
\begin{equation}
\label{intertwine}
\Z_8 \; :\;  \sigma \rightarrow \sigma+ \pi ~.
\end{equation}

The intertwining of the dual photon shift symmetry with continuous global symmetries, similar to our case of a discrete chiral symmetry (\ref{intertwine}), has been noted  in \cite{Affleck:1982as}.
Since (\ref{intertwine})  is not a continuous symmetry for $\sigma$,  but just a  $\Z_2$ shift symmetry, 
one may expect that an operator of the form $\cos (q \sigma), q=0({\rm mod} \; 2)$, may be induced.\footnote{ 
The BPS  and KK instanton amplitudes, and the 4d instanton 't Hooft vertices actually coincide  in 
 $\N=2$ SYM and the non-supersymmetric  QCD(adj) with two massless Weyl adjoint fermion. Consequently,  these two theories possess identical  discrete and continuous  chiral symmetries.   On the other hand, the magnetic bion operator is generated only in  SYM with $\N=1$ and  in   
 $\N=0$ QCD(adj). We expect that the microscopic reason behind the   non-formation of the magnetic bions in  $\N=2$ SYM  is  the existence of exactly massless adjoint Higgs scalars. 
  It is desirable to show this explicitly, and we hope to address this question in future work.   When the adjoint Higgs scalar is lifted by a soft mass term, the theory reduces to  $\N=1$ SYM, and a magnetic bion induced potential  $\sim \cos 2 \sigma$ is both permitted and generated.  }   However, a bosonic potential is forbidden by the large amount  of supersymmetry of the $\N=2$ theory, hence it is not generated.
In other words, since the  operators (\ref{2amplitudes}) in effective Lagrangian  have more than two zero modes, the $\N=2$ theory on   $\S^1 \times \R^3$   cannot  induce a superpotential and the moduli space is not lifted (in the  $\N=2$  theory,  the operators with four-fermion zero modes  contribute to the hyper-K\" ahler metric). The   IR physics is described  as a three-dimensional non-linear sigma model with target space ${\cal M}$, a hyper-K\" ahler manifold with quaternionic dimension one \cite{Seiberg:1996nz,Dorey:1997ij}.  The IR field theory is described in terms of gapless bosonic degrees of freedom, $\phi, \phi^{\dagger}, b \pm  i \sigma$, see (\ref{notation3}), and their fermionic superpartners. 

\subsubsection{$\mathbf{\N=1}$ perturbation at small  $\mathbf{\S^1 \times \R^3}$}
\label{sec:massatsmallL}

Adding a mass term for the chiral multiplet reduces the    $\N=2$ supersymmetry  to  $\N=1$ and  has the effect of lifting the $\psi$ zero modes from the instanton amplitudes 
(\ref{2amplitudes}). The mass perturbation (the kinetic term has an $\N=2$ normalization) is:
\begin{equation}
\int d^2 \theta\;  m  \; \tr \Phi^2 + \int d^2 \bar \theta \;  m \;  \tr \overline  \Phi^2  =  
m \; \tr (\bar \psi \bar \psi +  \psi  \psi)  + \ldots
\end{equation} 
Soaking up the $\psi$ zero modes from (\ref{2amplitudes}) with the mass perturbation induces the following 3d  
instanton amplitudes: 
\begin{eqnarray}
&&{\cal M}_{1} = m  e^{ - b +i \sigma} 
 \lambda \lambda, \qquad 
{\cal M}_{2} =m \eta  e^{ + b -i \sigma} 
 \lambda \lambda,  \cr  \cr
&&\overline {\cal M}_{1} = m e^{ - b - i \sigma} 
\bar \lambda \bar \lambda, \qquad 
\overline {\cal M}_{2} = m \eta  e^{ + b + i \sigma}
 \bar \lambda \bar \lambda, 
\label{1amplitudes}
\end{eqnarray}
each of which carries only two zero modes and, as in (\ref{2amplitudes}), we use $
\eta 
\equiv e^{2 \pi i \tau} =   e^{- \frac{8\pi^2}{g_4^2} + i \theta}$. 

  Four of the zero modes of the 4d instanton  are also lifted, as is transparent from the  combination ${\cal M}_{1} {\cal M}_{2} \sim \eta (\lambda \lambda)^2$.   
Consequently, the axial symmetry is reduced to $\Z_4$.   As in the $\N=2$ theory, the invariance of the amplitudes   (\ref{1amplitudes}) demands that $ \sigma \rightarrow \sigma+ \pi $ when we apply a discrete chiral rotation to fermions. Since (\ref{1amplitudes})  carry just two zero modes, they now also  generate  a superpotential, given by: 
 \begin{equation}
 W_{\R^3 \times \S^1}  \sim   m ( e^{- {\bf B}}
  +  \eta e^{\bf B} )~, 
    \label{2vac1}
 \end{equation} 
 where $\bf B$ is an $\N=1$ chiral superfield whose lowest component is $b - i \sigma$.
 Clearly,  there are two isolated vacua located located at:
 \begin{equation}
\langle  e^{\bf B}  \rangle =\langle  e^{b -i \sigma}  \rangle =  \pm \eta^{-{1\over 2}} = \pm e^{ {4 \pi^2 \over g_4^2} - i {\theta \over 2}}, ~{\rm or}\;\; \;\;  \langle a_4 \rangle = {\pi\over L},~~\langle \sigma \rangle= \left({\theta \over 2},\; {\theta\over 2} + \pi\right)~.
\label{minimum}
 \end{equation} 
Note that   (\ref{minimum}) implies that at $L \ll \Lambda_{\N=2}^{-1}$ the semiclassical reasoning is justified, as the vacuum is at the center-symmetric point $a_4 = {\pi\over L}$. 
 These results are well known \cite{Seiberg:1996nz,Aharony:1997bx,Davies:1999uw}. 
 
 Here, we would like to instead discuss the physics of the  superpotential in some detail. 
 Let us first study the bosonic potential,\footnote{\label{kinetic}The  kinetic terms of the fields $\sigma$ and $b$ are ${1 \over 2} {g_4^2\over (4 \pi)^2 L}\left[(\partial_i b)^2 + (\partial_i \sigma)^2\right]$, corresponding to a K\" ahler potential $K = {g_4^2\over 2 (4 \pi)^2 L} {\mathbf{B^\dagger B}}$.} omitting the inessential overall constant and taking $\theta = 0$:
  \begin{eqnarray}
V (b, \sigma) \sim   \left|\frac{\partial W}{\partial  {\bf B}} \right|^2  =   
e^{- 2b} + \eta^2  e^{ 2b} - \eta (e^{- 2i \sigma }  +  e^{ 2i \sigma }  )   ~,
  \label{bosonic}
\end{eqnarray}
which, when expanded around center-symmetric vacuum  $\langle b\rangle = {4 \pi^2 \over g_4^2}$,  takes the form:  
\begin{eqnarray}
V ( \langle b\rangle + b, \sigma) \sim   
\eta (e^{- 2b} + e^{ 2b} - e^{- 2i \sigma }  -  e^{ 2i \sigma }  ) \sim \eta \cosh 2 b - \eta \cos 2 \sigma ~. 
  \label{bosonic2}
\end{eqnarray}
Evidently, in the effective lagrangian, the mass gap for gauge fluctuations (recall that $\sigma$ is the dual photon) is generated by 
the operator  $e^{ \pm  2i \sigma }$, and for the spin-zero scalar  fluctuation, it is generated by  
$e^{ \pm  2b}$.   

As discussed in Ref.~\cite{Unsal:2007vu, Unsal:2007jx},  in gauge theories with massless adjoint fermions, the 3d instanton and twisted instanton do not  generate a mass gap for the gauge fluctuations due to their fermionic zero mode structure, dictated by the index theorem \cite{Nye:2000eg, Poppitz:2008hr}.
  Rather, the bosonic potential  is induced by monopole-antimonopole pairs (multi-instanton amplitudes), which may be viewed as composites, with opposite chirality zero modes soaked-up. 
The bosonic potential is sourced by the  amplitudes   $[{\cal M}_{1} \overline {\cal M}_{1} ]$, $[ {\cal M}_{2} \overline {\cal M}_{2} ]$, $[ {\cal M}_{1} \overline {\cal M}_{2} ]$, and $[ {\cal M}_{2} \overline {\cal M}_{1} ]$, which are composites of  (\ref{1amplitudes}). The magnetic and topological  charges  and the amplitudes    associated with these instanton-antiinstanton events are: 
   \begin{eqnarray}
      {\rm  composite}  \qquad &  \left(Q_m,  Q_T \right) \qquad  &   {\rm 
      amplitude}  \qquad   \;\;      \left( Q_{\rm dil} , \; Q_m,  \; Q_T \right) \qquad     \cr   \cr
  [{\cal M}_{1} \overline {\cal M}_{1} ]  \qquad  &   \; \; \; (0, 0)  \qquad   & \; \; e^{-2b}  \qquad    \qquad    \qquad   
   \;  (-2, \; \; 0, 0)  \qquad    \cr  \cr
 [ {\cal M}_{2} \overline {\cal M}_{2} ]  \qquad   & \; \; \; (0, 0)  \qquad   &  \; \; e^{+2b} \qquad  \qquad   \qquad
  \;    (+2, \; \; 0, 0)  \qquad    \cr \cr
  [ {\cal M}_{1} \overline {\cal M}_{2} ] \qquad  &  (+2, 0 ) \qquad   &  \; \;e^{+2i \sigma} \qquad  \qquad         \qquad
   (\;\;  0, +2, 0)  \qquad   \cr \cr
   [ {\cal M}_{2} \overline {\cal M}_{1} ] \qquad   & (-2, 0)  \qquad   &  \; \;e^{-2i \sigma}  \qquad    \qquad            \qquad
   (\; \; 0,-2, 0)  \qquad ~,
   \label{composites} 
\end{eqnarray} 
where the quantum numbers of the individual instantons are given in  (\ref{1amplitudes}). The last column and $Q_{\rm dil}$, a pseudo-quantum number,  will be explained in  Section 
\ref{sec:centersymmetry}. 

Eqn.~(\ref{composites}) listing the origin of the various terms in the scalar potential (\ref{bosonic}) is quite interesting. Note that all  $[{\cal M}_{i} \overline {\cal M}_{j} ]$ events have  vanishing topological charge,  i.e., they are indistinguishable from the perturbative vacuum in that sense. 
However, the $[{\cal M}_{1} \overline {\cal M}_{2} ]$   (and its anti-molecule) events carry two units of magnetic charge. Thus, they can  be distinguished from the vacuum and have been called ``magnetic bions" in \cite{Unsal:2007vu, Unsal:2007jx}.  They provide an example of stable semiclassically calculable bound states of a monopole-instanton and a twisted monopole anti-instanton. The contribution of the magnetic bion amplitude to the effective Lagrangian, adapting the results of  \cite{Anber:2011de,Unsal:2007jx} to the mass-deformed  $\N=2$ theory,  is: 
  \begin{eqnarray} 
&& [{\cal M}_{1} \overline {\cal M}_{2} ] \sim 
 {\cal A} e^{-2S_0} e^{2i \sigma},  \qquad {\rm where}  \cr \cr
&& {\cal A} \sim  \int_{0}^{\infty} dr e^{-V_{\rm eff, {\cal M}_{1} \overline {\cal M}_{2}}(r)} = \int_{0}^{\infty} dr e^{-  \left(2 \times \frac{4 \pi L}{g_4^2 r} + (4 n_f - 2) \log r \right)} ~, ~~n_f = 1~.
\label{m-bion}
\end{eqnarray}
 The physics behind the ``effective potential"  $V_{\rm eff}$ is that the ``magnetic" (due to exchange of $\sigma$) and ``electric" (due to exchange of $b$-scalar) repulsive interactions between the constituents of the bion\footnote{The fact that the $b$-field is not gapped (classically and to all orders in perturbation theory) in the supersymmetric case accounts the factor of two in front of the $1/r$ repulsion in $V_{\rm eff}$ compared to refs.~\cite{Anber:2011de,Unsal:2007jx}.} 
 are balanced by fermion zero-mode exchange  induced attraction (the choice $n_f=1$ reflects the presence of a single massless Weyl adjoint flavor in the mass-perturbed $\N=2$ theory). This results in stable topological molecules of size 
 $\ell_{\rm bion}
 \sim L/g_4^2$ (see \cite{Anber:2011de} for details) and a well behaved integral dominated around $r \sim \ell_{\rm bion}$.  
  The integral in the second line of (\ref{m-bion})  is over the radial separation 
 $|{\bf r}|= r $ between ${\cal M}_{1}$ and $\overline {\cal M}_{2} $ monopole-instantons in the Euclidean setting, and is an example  what is called a quasi-zero mode (the center of mass position of the $[{\cal M}_{1} \overline {\cal M}_{2}]$ molecule is an exact zero mode and the Gaussian fluctuations around these molecules are small). A correct treatment of the instanton-anti-instanton requires care due to these (quasi)-zero modes, and doing so yields a result in exact agreement with supersymmetry. In Section \ref{sec:centersymmetry}, we will identify a new type of topological molecule, which is more subtle to identify, but plays a useful role in center-symmetry realization. 
 
  It is important to note that the existence of these non (anti)self-dual topological excitations transcends supersymmetry: while studying only the superpotential (\ref{2vac1}) and inferring the resulting potential (\ref{bosonic}) correctly incorporates their effect by the familiar ``power of supersymmetry," it hides the physics of balancing repulsive and attractive forces, which also holds in  nonsupersymmetric theories with multiple adjoint fermions. 
 
Thus, we learn that in $\N=2$ SYM theory softly broken down to   $\N=1$, at small-$L$, the mass for the dual photon 
is induced by composite  topological excitations, the ``magnetic bions"
$ [ {\cal M}_{1} \overline {\cal M}_{2} ]$.\footnote{We note that the magnetic bion size 
in $\N=2$ broken down to $\N=1$ or $\N=0$ 
also  depends on the mass of the adjoint  Higgs scalar if the mass $m\lesssim L/g_4^2$. See 
Section~\ref{conclusions} for further discussion.} Clearly, this is  an exotic  generalization of Polyakov's confinement mechanism to a locally four 
dimensional gauge theory. In a straightforward generalization of Polyakov's mechanism, the mass gap would occur due to operators  $e^{-S_0} e^{\pm   i \sigma } $, typical for monopole-instantons 
without any fermion zero modes.  This operator is forbidden in our theory 
due to the discrete chiral symmetry (\ref{intertwine}), but the operator $e^{2 i \sigma}$ is allowed and generated by magnetic bions. 

It is well known that the generation of mass gap for the dual photon $\sigma$ also implies confinement of electric charge \cite{Polyakov:1976fu}. The mass gap for gauge fluctuations (the mass, $M$, of $\sigma$ and  $b$) and the  tension of the confining electric flux tube $\Sigma$  between two static sources can be semiclassically calculated:
\begin{eqnarray}
M^2 &\sim&  {m^2 \over g_4^8} \; e^{- {8 \pi^2 \over g_4^2}} \sim m^2 \times (\Lambda_{\N=2} L)^4, \nonumber \\
\Sigma &\sim& {g_4^2 \over L } \;  M \sim m \Lambda_{\N=2}  \times (\Lambda_{\N=2} L)~,
\label{tensionsmallL}
\end{eqnarray}
where the coupling is taken at the scale $L$ and subleading logarithmic dependence on $\Lambda_{\N=2}L$ is neglected in the last equality on each line. The mass gap $M$, including the $g_4$-dependent prefactor can be inferred  from the recent calculation in multi-flavor QCD(adj) of \cite{Anber:2011de}, while the result for the string tension follows from \cite{Polyakov:1976fu}.\footnote{The same $g_4^2$ scaling of the mass gap as in (\ref{tensionsmallL})  is obtained if one follows the normalizations in the $\N=1$ supersymmetric calculation of  \cite{Davies:2000nw}.} 

Equations (\ref{tensionsmallL}) for the mass gap and string tensions can be compared to the corresponding expressions for $\R^4$ given in   (\ref{swstringtension}). Clearly the mass gap  $M^2$ in (\ref{tensionsmallL}) and on $\R^4$ have different power dependence on $m$, while the string tensions $\Sigma$ at both large and small $L$ scale as $m \Lambda_{\N=2}$. The small-$L$ values of $\Sigma$ and $M$ both increase with $L$ at fixed $\Lambda_{\N=2}$ and presumably saturate to the $\R^4$ values near $\Lambda_{\N=2}L \sim 1$, the region where we lack control over the theory (the string tension $\Sigma$ can be defined at any $L$ from the area law obeyed by a Wilson loop of appropriate size  $\in \R^3$ and $M$ from the exponential fall-off of the two-point function of the gauge invariant magnetic field strength $B^k \sim \epsilon^{kij} \tr\left[F_{ij} (\Phi + {1 \over L}\Omega) \right]$, where $\Omega$ is the Wilson line around $\S^1$).

As opposed to magnetic bions, the other composites from (\ref{composites}), the   $[{\cal M}_{i} \overline {\cal M}_{i} ]$ amplitudes, responsible for generating a potential for $b$ (or $a_4$),  do not even carry a magnetic charge!   Yet, both types of topological defects  play  crucial  role in the dynamics, including center-symmetry realization and confinement, as we now discuss.

 \subsubsection{Center-symmetry realization and  center-stabilizing bions}
 \label{sec:centersymmetry}
 The potential for the $b$-field,  $\sim e^{- 2b} + \eta^2  e^{ 2b} 
$, from eqn.~(\ref{bosonic}) generates a non-perturbatively induced repulsive interaction between the eigenvalues of the Wilson line around $\S^1$.  The minimum (\ref{minimum}) is at $b=  \frac{4 \pi^2}{g_4^2} $, or, in terms of the Wilson line, it is:
\begin{equation}
\Omega= e^{i A_4 L} = 
\left( \begin{array}{cc}   e^{i \frac{\pi}{2}} &  \\ & e^{-i \frac{\pi}{2}}  \end{array}    \right) , \qquad 
\tr \Omega =0~,
  \label{Wilsonline} 
\end{equation} 
up to gauge rotations. This is the unique center-symmetric vacuum of the theory on $\R^3 \times \S^1$. 
 The origin of the center-stabilizing operator in the Lagrangian  are the $[{\cal M}_{1} \overline {\cal M}_{1} ]$ and $[{\cal M}_{2} \overline {\cal M}_{2} ]$ induced amplitudes  in (\ref{composites}).\footnote{The remarks here and in Section \ref{sec:chiralsymmetry}  below hold for the mass-deformed $\N=2$ theory as well as for the pure $\N=1$ super-Yang-Mills theory, with the appropriate scale matching $\Lambda_{\N=1}^3 = m \Lambda_{\N=2}^2$ applied; for this  reason, in eqns.~(\ref{fermionmass}), (\ref{runaway}), and (\ref{4dW}) below we omit the factors of $m$ (along with other dimensionful factors) in the superpotential. }  

 The amplitude associated  with $[{\cal M}_{1} \overline {\cal M}_{1} ]$ generates the run-away potential $e^{- 2b}$ for the eigenvalues of the Wilson line and forces them to be as far apart as possible. This part is similar to the lifting of Coulomb branch in  $\N=2$ SYM on $\R^3$,  where the quantum theory does not have a ground state (or it is pushed to $b=\infty$) \cite{Affleck:1982as}.    However, our theory has two interrelated differences with respect to  $\N=2$ SYM on $\R^3$. The classical moduli  is compact and the theory has an extra set of topological molecules,  $[{\cal M}_{2} \overline {\cal M}_{2} ]$.   Were it not for  $[{\cal M}_{2} \overline {\cal M}_{2} ]$, the two eigenvalues would end up at $\pi$, corresponding to a center broken Wilson line, $\Omega= -1$. However, quite symmetrically,  
 $[{\cal M}_{2} \overline {\cal M}_{2} ]$ generates a repulsion between the two eigenvalues of the Wilson line which prevents them from coinciding. Consequently, the combination of the two center-stabilizing bions is to yield the center-symmetric minimum   (\ref{Wilsonline}.)  
 
Here, we give a brief description of how the $[{\cal M}_{1} \overline {\cal M}_{1} ]$ and $[{\cal M}_{2} \overline {\cal M}_{2} ]$ contributions arise.  A detailed discussion of these type of topological molecules in  both supersymmetric and non-supersymmetric gauge theories will appear in \cite{Argyres:2011xx} and  the implication for the thermal deconfinement phase transition will appear in  \cite{Poppitz:2011xx}.
The contribution of the   $[{\cal M}_{i} \overline {\cal M}_{i} ]$ amplitude to the effective theory 
is, naively,  
  \begin{eqnarray} 
&& [{\cal M}_{1} \overline {\cal M}_{1} ] \sim 
 {\cal A} e^{-2S_0} e^{\pm 2b } , \qquad {\rm where}  \cr \cr
&& {\cal A} \sim  \int_{0}^{\infty} dr e^{-V_{\rm eff, {\cal M}_{1} \overline {\cal M}_{1}}(r)} = \int_{0}^{\infty} dr e^{-  \left(-2 \times \frac{4 \pi L}{g_4^2 r} + (4n_f -2) \log r \right)} ~,  \; n_f=1~,
\label{e-bion}
\end{eqnarray}
 where  now the interaction  between constituents is   all attractive: 
  ``magnetic" (due to exchange of $\sigma$-scalar) and ``electric" (due to exchange of $b$-scalar) attractions  and the  fermion  induced attraction (again, we put $n_f=1$ for the case of interest).  The integral is dominated 
  by the small-$r$ domain, where not only (\ref{e-bion}) is incorrect, it is also hard to make sense of constituents as the interaction becomes large.  This is in sharp contrast with the magnetic bion  \cite{Unsal:2007jx,Anber:2011de}.  

 In contrast with the magnetic bions, these instanton--anti-instanton ``molecules" are difficult to exhibit semiclassically, as their constituents have only attractive interactions,  and  naively the natural tendency is for these objects to annihilate.  However, the amplitude associated with 
  $[{\cal M}_{i} \overline {\cal M}_{i} ]$ is not proportional to identity operator, but rather  to $
  e^{\pm  2b}$.  This means that, although in the sense of  magnetic and topological charge these defects are indistinguishable from the perturbative vacuum,  since the product of the amplitudes   
 $[{\cal M}_{i} \overline {\cal M}_{i} ]$  cannot be contracted to the identity, it should perhaps be seen as carrying a  {\it pseudo-quantum number}  to distinguish it from perturbative vacuum. In fact, the coupling of $b$ to monopole-instantons can be thought of as the coupling of a massless ``dilaton"   \cite{Harvey:1996ur} (dilatation is a classical symmetry, only globally broken  by compactification and locally by the expectation value of $b$) or as ``electric" charge, if one thinks of the compact $x_4$ as Euclidean time. Since the adjoint Higgs field is asymptotically  
 of the form $A_4 = a_4 T^3 \left(1 - \frac{1}{a_4 r} + \ldots \right)$, we can define a ``dilaton" 
 charge (or flux)  associated with it by using Gauss' law:  
\begin{equation}
 Q_{\rm dil} = \int_{S^2_{\infty}} \vec \nabla A_4 \cdot d\vec S ~.
 \end{equation}
 Note that  $\vec \nabla A_4$ is the dimensional reduction of the {\it Euclidean} electric field ${\vec E}$ and integral may be  interpreted, in some loose sense, as  electric flux.\footnote{However, 
 this is rather confusing  as it forces us to think of monopole-instantons on $\R^3 \times S^1$ as having both real electric and magnetic charges and leads to possible confusion with magnetic dyon particles on $\R^4$ (which carry genuine electric and magnetic charges). In the literature, the latter terminology is (appropriately)  used in thermal Yang-Mills theory, see Ref.\cite{Diakonov:2009jq}, but here  we refer to $Q_{\rm dil}$  as dilaton charge to prevent confusion, and refer to ${{\cal M}_1}$ as monopole-instanton.}

 Therefore, a more pragmatic way to think of the quantum numbers for topological defects on $\R^3 \times \S^1$ is to incorporate   $Q_{\rm dil}$ and generalize the doublet of quantum numbers to a triplet: 
 \begin{equation}
 (Q_m, Q_T) \longrightarrow ( Q_{\rm dil}, Q_m, Q_T) 
 \end{equation} 
 In this language,  the charges of  ${\cal M}_{1}$ are $(-1, +1, \half)$ and the charges of 
  $[{\cal M}_{1} \overline {\cal M}_{1} ]$  are $(-2, 0, 0)$, as also shown in the last column in (\ref{composites}). In this sense,   $[{\cal M}_{1} \overline {\cal M}_{1} ]$  molecule is now distinguishable from perturbative vacuum. 
  Monopoles and anti-monopoles carry same  ``dilation" or ``electric" charges,  and opposite magnetic charges. Also note that the same dilaton charge objects {\it attract}, as opposed to 
  the fact that same magnetic charge objects repel.

Finally, we note that supersymmetry demands that the  operators induced by instanton-anti-instanton molecules  should be there and their coefficients are identical to the one of magnetic bion up to a sign, see (\ref{bosonic2}). 
In fact, using analytic continuation in the coupling constant \cite{Bogomolny:1980ur, ZinnJustin:1981dx} or in the 
 path integration contour over the quasi-zero mode \cite{Balitsky:1985in}, such saddle points of the Euclidean path integral can  be defined. In particular, in supersymmetric quantum mechanics  \cite{Balitsky:1985in} and even $\N=1$ supersymmetric QCD \cite{Yung:1989aa} these complex saddle point contributions to the Euclidean path integral seem to be unambiguously defined. 
Since they are required by supersymmetry, we suspect  that this is also the case in the mass-perturbed $\N=2$ theory (or in pure $\N=1$) at small $L$, an issue which is discussed in more detail  in  \cite{Argyres:2011xx, Poppitz:2011xx}.


 \subsubsection{Chiral symmetry and the topological disorder operator}
 \label{sec:chiralsymmetry}
 
 The potential for the dual photon field  $\sigma$ in (\ref{bosonic}) is $- \cos 2 \sigma$, with two isolated minima (\ref{minimum}) located (for $\theta =0$) at $\sigma=0$ and $\pi$. The two minima are related  to each other  by the exact $\Z_4$ chiral symmetry of the mass-deformed $\N=2$ theory. The  potential for the dual photon is generated  by magnetic bion ``molecules," $[{\cal M}_{1} \overline {\cal M}_{2} ]$, of magnetic charge $\pm 2$, whose dynamical stability is semiclassically calculable, as already explained.

The order parameter for chiral symmetry  in the small-$\S^1$ domain of mass deformed Seiberg-Witten theory (as well as in the pure $\N=1$ theory) is:
\begin{equation}
\label{disorderoperator}
\langle e^{i \sigma} \rangle = \pm 1. 
 \end{equation}
Contrary to assertions in literature stating that the chiral symmetry is broken 
by a local fermion bi-linear $\langle \tr \lambda  \lambda \rangle$ \cite{Davies:1999uw},  chiral symmetry is in fact broken by the vacuum expectation value of the topological disorder operator (\ref{disorderoperator}). 
If one performs a small-$L$ monopole-instanton calculation of the fermion bilinear expectation value  in the full $SU(2)$ gauge theory, $\langle \tr \lambda  \lambda  \rangle$, similar to   \cite{Davies:1999uw}  but   incorporating the long-range interactions in the monopole-antimonopole (or more precisely, bion-antibion) plasma, one finds that it is related  to the expectation value of the disorder operator (\ref{disorderoperator}):
\begin{equation}
\langle \tr \lambda \lambda \rangle \sim \langle e^{i \sigma} \rangle \eta^{1 \over 2}~.
\label{condensate}
\end{equation} 
This shows that the  source of chiral symmetry breaking is the magnetic-bion induced potential (\ref{bosonic})  for the dual photon. A further argument that (\ref{condensate}) is correct is that it correctly reproduces the   values for the gaugino condensate in the two vacua (\ref{minimum}), while due to the omission of the $\langle e^{i \sigma} \rangle$ factor on the r.h.s.,  Eqn.~(4.7) in \cite{Davies:1999uw} only gives a single value.\footnote{The breaking of chiral  discrete or continuous symmetries by the disorder operator expectation value is also a generic feature of  nonsupersymmetric gauge theories with fermions on $\R^3 \times \S^1$ \cite{Poppitz:2009tw, Poppitz:2009uq}.}
 
As usual, the spontaneous breaking of chiral symmetry generates  fermion mass. 
 Consider the fermion bilinear terms arising from the superpotential:
\begin{eqnarray}
\label{fermionmass}
 \frac{\partial^2 W}{\partial  B^2}  \; \lambda \lambda + {\rm c.c.}  \sim  
( e^{- b +i\sigma} + \eta e^{+ b -i\sigma}  )\;   \lambda \lambda + {\rm c.c.} ~,
\end{eqnarray}
and evaluate them at the center-symmetric vacuum (\ref{minimum}), to find that the fermion mass breaking the $\Z_4$ chiral symmetry:
\begin{eqnarray}
 \eta^{\half} (\langle e^{i \sigma}  \rangle + \langle e^{-i \sigma}  \rangle ) \;  \lambda \lambda = \pm 2 \eta^{1 \over 2} \;\lambda \lambda~,
\end{eqnarray}
is also due to (\ref{disorderoperator}). 


Finally, we can consider taking the small and infinite $L$ limits of the superpotential (\ref{2vac1}).
In the $\R^3$ limit,  keeping $g_3^2 = g_4^2 L$ fixed,  we find that $\eta \rightarrow 0$ 
 and the  two vacua (\ref{minimum})  run-away to infinity. The superpotential reduces to the well-known run-away superpotential: 
 \begin{equation}
 W_{\R^3 }   \sim   e^{- {\bf B}}~.
     \label{runaway}
 \end{equation}
  In the $\R^4$ limit, we should integrate out the chiral superfield $e^{- {\bf B}}$, as it does not represent a valid infrared degree of freedom---this can be seen, e.g., from the fact that the $\sigma,b$ kinetic terms, see Footnote~\ref{kinetic}, vanish in the infinite-$L$ limit; see also \cite{Aharony:1997bx}.  
   In doing so,  we obtain the four dimensional gaugino condensate superpotential:
  \begin{equation}
  \label{4dW}
 W_{\R^4}  \sim  \pm   \eta^{1/2}~,
 \end{equation}
corresponding to the two isolated vacua of the 4d  mass perturbed $\N=2$ or pure $\N=1$  theory.

\section{Phase diagram and abelian (non-'t Hooftian) large-N limits}
\label{phase}

On  $\R^4$,  confinement in an $SU(2)$ gauge theory with  $\N=2$ supersymmetry  softly broken   down to $\N=1$  $(m \ll\Lambda_{\N=2})$  is a version of  abelian confinement.  
By this we mean that the long-distance effective Lagrangian is an abelian  $U(1)$ gauge theory,  despite the fact that the microscopic theory is a non-abelian  $SU(2)$. The confinement of the electric charges is  due to magnetic monopole or dyon condensation \cite{Seiberg:1994rs}.
 
On the other hand, in the limit $m \gg \Lambda_{\N=2} $, where the adjoint Higgs multiplet decouples and the theory reduces to pure $\N=1$ SYM theory, there exists no known description of the gauge dynamics  on $\R^4$  where abelianization takes place. It is usually believed that there is no phase transition as the mass term is dialed from small to large and the theory moves from a regime of abelian confinement to non-abelian confinement.

In the {\it pure}  $\N=1$ SYM theory, as well as in a large-class of non-supersymmetric gauge theories which remain center-symmetric upon compactification down to small radius, it has been recently understood that 
the  $L  \Lambda_{\N=1} \ll 1   $ regime also exhibits abelian confinement.  The confinement of the electric charges is  now due to the magnetic bion mechanism \cite{Unsal:2007vu,Unsal:2007jx}.  Analogous to the mass-deformed theory, it  is usually believed that there is no phase transition associated with center symmetry as the radius   is dialed from small to large. At large-$L$,  it is expected that   non-abelian confinement should take place.

  \begin{figure}[h]
 \begin{center}
\includegraphics[angle=-90, width=6.0in]{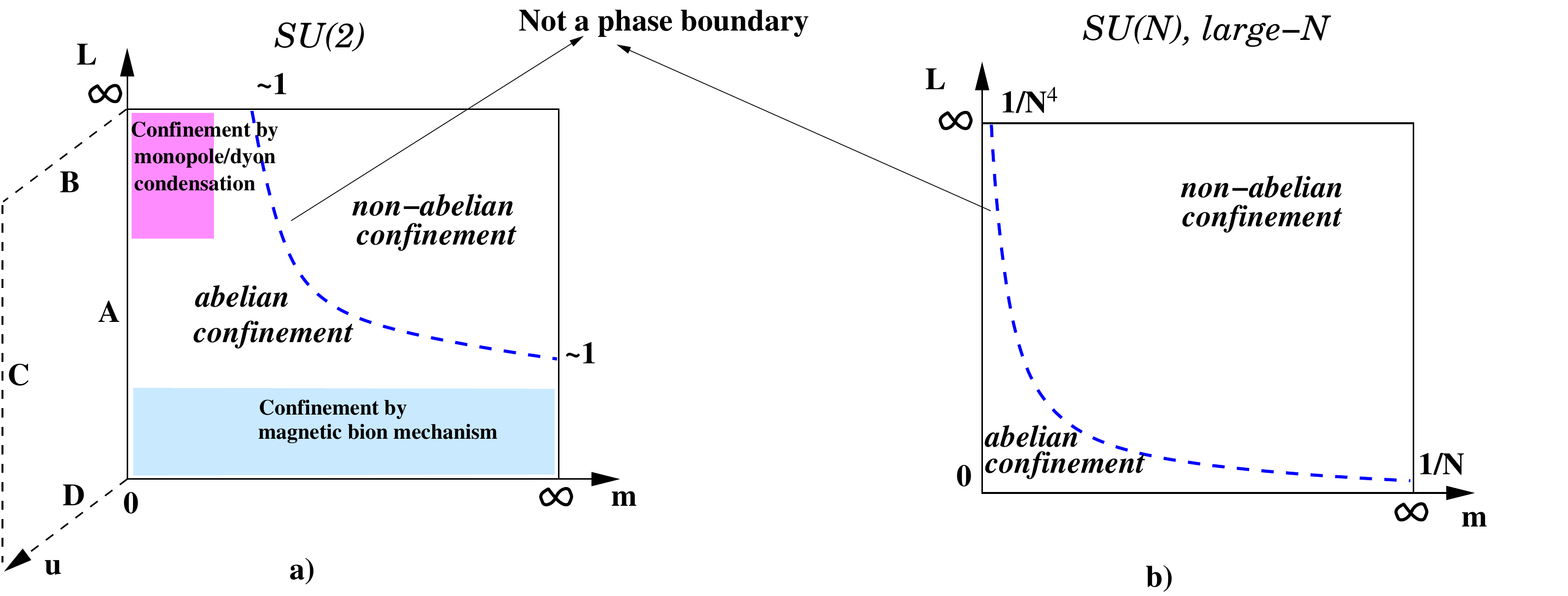}
\caption{ The $\N=2$ theory broken down to   $\N=1$ exhibits confinement.  At small $m$ 
and/or small $L$ (in units of $\Lambda$) the dynamics abelianizes at large distances and the theory exhibits 
abelian confinement.  The phase diagram in the $m$-$L$ plane for the small-$N$ theory is shown on the left figure, where the shaded areas indicate the calculable regimes at small and large $L$. The third ($u$) direction---which allows to smoothly connect the   topological excitations responsible for confinement at large and small $L$ via Poisson duality---is also indicated. 
At large-$N$, shown on the right figure, the calculable semi-classical confinement regime shrinks to a narrow sliver both in $m$ and $L$, in a correlated manner, as explained in the text. 
 }
  \label {fig:phase}
 \end{center}
 \end{figure}

 At what values of $m$  and $L$ does the metamorphosis from the abelian confinement (which we analytically understand) to the non-abelian confinement  (which is not yet understood) take  place?  Naively, by dimensional analysis, one may argue that this should happen around 
$m \sim \Lambda_{\N=2}$  for the theory on $\R^4$ and  at $L \sim \Lambda_{\N=1}^{-1}$ for the pure $\N=1$ SYM.   Although this is true for $SU(N)$ gauge theory with $N=2$, or a few, as shown in Fig.~\ref{fig:phase}.a), it turns out to be incorrect 
especially at larger values of $N$. 

 This subtlety is associated with the  regimes of gauge theory  in which the long distance dynamics reduces to the one of  the abelian subgroup $U(1)^{N-1}$ and with the existence of light $W$-bosons whose masses scale as ${\cal{O}}(1/N)$. This has been previously discussed by Douglas and Shenker in the mass deformed $SU(N)$ $\N = 2$ supersymmetric Yang-Mills theory   for $m$-scaling  \cite{Douglas:1995nw} and by Yaffe and one of us (M.\"U.) in the context of  
 center symmetric vector-like gauge theories for  $L$-scaling   \cite{Unsal:2008ch}.
 
In this work,  we  argue that these two large-$N$ limits are indeed correlated in the phase diagram of the  theory in $L$-$m$ plane.  

At large $\S^1$, in the  $N\rightarrow \infty $ limit of the mass deformed theory, the Abelian long distance regime is preserved only if the mass deformation\footnote{Note that at large-$N$ the mass deformation is $W = N m \tr \Phi^2$ \cite{Douglas:1995nw}.} $m$ is sent to zero as  $m   \sim \Lambda_{\N=2}/N^4$ (while the naive expectation would have been that $m \ll \Lambda_{\N=2}$ suffices). This follows from demanding the string tensions (see (\ref{swstringtension}), which remains valid at large $N$ for the lowest string tensions) to be smaller than the lightest $W$ boson mass squared, $m_W \sim \Lambda_{\N=2}/N^2$  \cite{Douglas:1995nw}, 
 a requirement  which is crucial for the validity of the abelianized effective theory. Equivalently, 
 the abelianized  low energy effective theory is valid provided there exists a hierarchy of scales between the heaviest U(1)-photons and lightest W-bosons, which translates into: 
 \begin{equation}
 \frac{m_\gamma}{m_W}
  \sim  \sqrt \frac{ m}{ \Lambda_{\N=2}}N^2 \lesssim 1  \qquad (L=\infty)~.
\end{equation}

At small $\S^1$, on the other hand, it is the radius $L$ that has to be taken to zero in order for the abelianized description to be valid as $N\rightarrow \infty$. To see this, first 
take the large-$m$ limit with the appropriate scale matching    $\Lambda_{\N=1}^3=  m  \Lambda_{\N=2}^2$. Then,
 we find that the abelian long distance regime is preserved only if the compactification radius  $L$ is sent to 
 zero as $L \sim 1/(\Lambda_{\N=1} N)$. This follows from demanding that the mass of the lightest $W$ boson,  $m_W \sim 1/(LN)$ in the center-symmetric vacuum, be  parametrically larger than the heaviest dual photon mass  $m_\gamma$; we find: 
  \begin{equation}
 \frac{m_\gamma}{m_W}
  \sim  (L \Lambda_{\N=1} N)^3 \lesssim 1  \qquad (m=\infty)~.
\end{equation}
 
 The conditions used in  Ref.\cite{Douglas:1995nw} and  Ref.\cite{Unsal:2008ch}  for the  validity of an abelianized   low energy description  are in fact equivalent. 
Thus, connecting these two large-$N$ limits, we conjecture  that an  abelian  long distance regime only survives in a  corner which becomes arbitrarily narrow  as $N \rightarrow \infty$, as shown in Fig.~\ref{fig:phase}.b). In particular, at any fixed non-zero  $L\sim O(N^0)$ and $m\sim O(N^0)$,  if one takes $N\rightarrow \infty $ first,   there is no regime of the supersymmetric gauge theory in which the long distance dynamics remains abelian. At $L\sim O(N^0)$  and $N\rightarrow \infty$, the theory exhibit volume independence, and  behaves as if it is 
decompactified even when $L \Lambda_{\N=1} \ll 1$ ,  see for example \cite{Unsal:2008ch}.
 At any 
$m/ \Lambda_{\N=2} \sim O(N^0)$, as  $N\rightarrow \infty$, we expect the adjoint scalar to completely decouple  even when $m/ \Lambda_{\N=2} \ll 1$, and the $\N=1$ dynamics to be independent of $m$. 
This mass independence is the equivalent of the large-$N$ volume independence in the non-abelian confinement domain.  In other words, in the large-$N$ limit, we conjecture that all points in the non-abelian confinement domain in Fig.~\ref{fig:phase}.b) are equivalent up to subleading corrections in $N$.

\section{Discussion  and open problems} 
\label{conclusions}

By using field theory techniques 
and with rather  minimal help from supersymmetry, we were able to  answer the physical questions that we posed in the Introduction.  For the $\N=1$ mass deformation of the $\N=2$ theory, we have shown that:
  \begin{itemize}
 \item[{\it i)}]  The theory confines and exhibits a mass gap through the magnetic bion mechanism. This mechanism also applies to a large class of non-supersymmetric theories, notably to QCD with multiple massless adjoint Weyl fermions. 
   \item[{\it ii)}]  The discrete chiral symmetry is broken due to the condensation of a disorder operator, and the $SU(2)$ theory has two isolated vacua.  In the effective long distance theory,  this is the source of dynamical mass generation for fermions. The magnetic bions  responsible for chiral symmetry breaking  are the ${\cal M}_1  \overline {\cal M}_2$  (3d BPS instanton-twisted-antiinstanton)   and  ${\cal M}_2  \overline {\cal M}_1$  (3d BPS anti-instanton-twisted instanton) topological molecules, generating the operators $e^{\pm i 2 \sigma}$. These objects carry magnetic charge $Q_m = \pm 2$ and their stability can be semiclasically understood.
\item[{\it iii)}]  The center symmetry is stabilized through non-perturbative  ${\cal M}_1  \overline {\cal M}_1$  (3d BPS instanton-antiinstanton)   and  ${\cal M}_2  \overline {\cal M}_2$  (twisted instanton-antiinstanton)  topological molecules generating the operators $e^{\pm 2 b}$. These objects differ  from the perturbative vacuum because they carry $Q_{\rm dil}= \pm 2$ units of ``dilatonic" charge.

\item[{\it iv)}] The Seiberg-Witten   \cite{Seiberg:1994rs} and Polyakov \cite{Polyakov:1976fu}  solutions on  
$\R^4$ and $\R^3$  have been  known for almost two  and over three decades, respectively.   Recent  progress in understanding confinement in gauge theories on $\R^3 \times \S^1$,  in particular   the Polyakov-like magnetic bion mechanism in QCD(adj) and $\N=1$ SYM \cite{Unsal:2007vu, Unsal:2007jx},  permits us to demonstrate that the  two type of confinements are in fact continuously connected.  Namely, the topological excitations responsible for confinement at small- and large-$L$: the winding monopole-instanton constituents of the magnetic bions and the dyon particles, respectively, are related by Poisson resummation.  
 
  \end{itemize}

{\flushleft{There is a number of interesting questions  that arise related to our construction:} }

\begin{enumerate}
\item {\it  Non-formation of magnetic bions in  $\N=2 $  theory:} In the pure $\N=2$ theory, the global symmetries (except supersymmetry)  are identical to those of non-supersymmetric QCD with $n_f=2$ flavors of Weyl adjoint fermions.  The latter theory,  as well as the supersymmetric  $\N=1$ theory, permit the  magnetic bion induced operator  $\sim \cos 2 \sigma$, but the theory with  $\N=2$ supersymmetry does not. 
  The fact that supersymmetry does not permit this operator is clear. Since leading topological defects have four zero modes, see (\ref{2amplitudes}), they do not induce a superpotential hence a bosonic potential is also not induced.
  However, while true as a symmetry argument,  we think that it is interesting to understand, by microscopic means,  the   non-formation of  magnetic bions in  $\N=2$ SYM.  We believe the non-formation of magnetic bions in  $\N=2 $  theory is  due to the existence of  exactly massless adjoint Higgs scalars counter-acting the multi-fermion induced attraction and that it would be of interest to show this explicitly. 
  
\item  {\it  The size of magnetic bions in softly broken $\N=2 $  theory at small $m$:}   
 In the pure $\N=1$ theory, which is the $m=
\infty$ limit of deformed SW theory with appropriate coupling matching, the typical magnetic bion size is---in a regime where we have semi-classical control over the dynamics---${\ell}_{\rm bion}(m=\infty) \sim  L/g_4^2$.   In the $\N=2$ theory with mass deformation, we expect the bion size to be dependent on $m$.  If the correlation length for the scalar is smaller than the bion size, i.e., $m^{-1} < {\ell}_{\rm bion}(\infty) $, the scalar decouples quickly and the bion size should remain as it is in the $\N=1$ theory. However, if 
  $m^{-1} < {\ell}_{\rm bion}({\infty}) $, then we expect that the bion size should be proportional to 
  an inverse power of $m$   and diverge in the $m=0$ limit. This is necessary for the non-formation of the bions in the pure $\N=2$ theory.   A different behavior would invalidate the magnetic bion description of confinement in softly broken $\N=1$. Thus, it is desirable to study the adjoint mass dependence of the magnetic bion size.  
  
 \item {\it Generalizations:}  It would be of interest to generalize our discussion to other gauge groups and consider the inclusion of other  matter representations, where, possibly, new phenomena may be observed.

 \item {\it Topological molecules and  the deconfinement transition:} We have identified 
 the topological  molecules whose main role is to provide a center-symmetric vacuum. Similar and related molecules are also present in non-supersymmetric gauge theories, 
 including pure Yang-Mills. These defects may play the pivotal role in the 
 deconfinement phase transition, and may lead us towards  a microscopic theory of deconfinement. Work in this direction is ongoing.

\item {\it Relation to lattice studies of confinement:}
Let us finally comment on the non-supersymmetric case. The bion confinement mechanism operative at small $L$ in the mass deformed $\N=2$ theory also applies to a large class of non-supersymmetric theories, notably to QCD(adj) theories with multiple massless adjoint Weyl fermions. This class of non-supersymmetric theories provides the first example where confinement and chiral symmetry breaking can be studied in a controlled manner in a locally 4d theory. 
 
It would be very interesting to know whether, in the 
non-supersymmetric  multiple-adjoint fermion ``QCD-like" theories,  a   relation can be inferred between the confinement mechanism at small $L$ (which is reliably described by the magnetic bion mechanism) and large $L$ (where it is not understood). In the large-$L$ limit,  lattice studies of pure Yang-Mills theory  have shown the relevance of topological defects charged under the Abelian part (or the center) of the gauge group to the generation of mass gap and confinement (for a recent review, see~\cite{Greensite:2011zz}). Needless to say, this problem---addressing which would likely require both analytic and lattice input---is left for future studies.
  \end{enumerate}

 \acknowledgments We  thank P. Argyres, M. Shifman,  A. Vainshtein, and A. Yung   for useful discussions.   This work was supported by the U.S.\ Department of Energy Grant DE-AC02-76SF00515 and by the National Science and Engineering Council of Canada (NSERC).

\appendix 

\section{Chirality and fermion zero modes}
\label{chiralityappendix}

In the presence of massless fermions, the instanton amplitudes on $\R^4$ as well as the 
3d instantons (either at large or small $L$) acquire fermion zero modes.   The number of zero modes is dictated by various index theorems, APS on $\R^4$, Callias on  $\R^3$, and Nye-Singer on $\R^3 \times \S^1$ interpolating between the two. 
In our discussion of Poisson duality, we have  asserted certain conditions about  the fermion 
zero modes and  preserved supersymmetries in the background of topological defects. In particular,  we will show that the duality described in    (\ref{Poissond})  is unharmed by the 
inclusion of the four-fermi operators on both sides.

The issues about chirality and zero modes may be succinctly  described  
starting with  the supersymmetry transformation properties of a six dimensional theory, and then applying dimensional reduction in 
$x^{5,6}$ direction and compactification in $x^4$ direction. The notation is given in 
Section~\ref{sec:6d}.

\subsection{4d-instanton}
  As a warm-up, let us start with a 4d instanton. The classical instanton configuration is the solution of  self-duality condition  $F_{\mu \nu} = \half \epsilon_{\mu \nu \rho \sigma} F^{\rho \sigma}$  supplemented with  vanishing fermionic fields.  
Since supersymmetry transformation  relates variation of bosonic fields to fermions,    the  variation of bosonic fields are all zero, and  the instanton  background preserves  the supersymmetries for which    variation of fermionic field vanishes, i.e., 
$\delta \Psi =0$. 
The complement, which is  not annihilated under supersymmetry transformation, 
  gives the zero mode solution to the Dirac equation under the background of the instanton (or relevant topological excitation.) 
      
In the background of an instanton, setting the $A_5$ and $A_6$ scalars to  zero, the supersymmetry variation of the fermion is ($\Gamma^{MN} = [\Gamma^M, \Gamma^N]/2$): 
  \begin{eqnarray}
  \label{4dinstsusy}
 \delta \Psi  =&&( \Gamma^{MN}F_{MN}  )  \; \varepsilon \cr \cr
 =&&  \Big( \Gamma^{\mu\nu}F_{\mu\nu} + 2 \sum_{m=5,6} \Gamma^{\mu m}D_{\mu}A_m +  i \sum_{m,n=5,6}   \Gamma^{mn}[A_{m}, A_n] 
                  \Big)   \varepsilon  \cr
   =&&  \left(  \half   \Gamma^{\mu\nu}F_{\mu\nu} + \half  \Gamma^{\mu\nu} ( \half  \epsilon_{\mu \nu \rho \sigma} F^{\rho \sigma})               \right)      \varepsilon              \cr    \cr
                         =&&  \half F_{\mu\nu}  \left( \Gamma^{\mu\nu} + \half \epsilon^{\mu \nu \rho \sigma}                            \Gamma_{\rho \sigma}   \right)      \varepsilon              \cr     \cr                      = &&  \left[  \half F_{12} ( \Gamma^{12} + \Gamma^{34}) +    \half F_{13} ( \Gamma^{13} - \Gamma^{24}) +                          
                         \ldots                  \right]   \varepsilon           ~,                                   
  \end{eqnarray} 
  where we used the self-duality of the solution in the third line above. 
The final equality in (\ref{4dinstsusy}) implies,  for non-vanishing $F_{\mu \nu}$, that   the supersymmetry preserved for a self-dual instanton is the one with parameter defined by the first line below, while a similar argument for the anti-self-dual instanton yields the supersymmetry defined by the second line:
    \begin{eqnarray} 
&& \Gamma^{1234}  \varepsilon =  + \varepsilon \qquad  { (\rm 4d-instanton \; \;  background) } \cr
 &&  \Gamma^{1234}  \varepsilon =  - \varepsilon \qquad  { (\rm 4d-anti-instanton \; \;  background) }~,
\end{eqnarray}
with $\Gamma^{1234} = \Gamma^1 \Gamma^2 \Gamma^3 \Gamma^4$.
This defines a chirality condition for the zero mode  structure of a 4d instanton.  We use a basis 
where  the instanton zero modes are chiral and anti-instanton zero modes are anti-chiral.

 \subsection{3d monopole-instanton tower}
Now, consider the first order differential equations for the monopole-instantons that we derived in Section \ref{sec:bos} and figure out which half of the supersymmetries are preserved in this background for a given value of   
 $\alpha$. The preserved supersymmetries can be found analogously, using $A_6 = [A_4,A_5] = 0$ and eqn.~(\ref{second}) for the BPS monopole-instantons with $B_i ={1\over 2} \epsilon_{ijk} F_{jk}$: 
 \begin{eqnarray}
 \delta \Psi= &&( \Gamma^{MN}F_{MN}  )  \; \varepsilon \cr \cr
                =&&  \left( \Gamma^{ij}F_{ij} + 2 \sum\limits_{m=4,5,6} \Gamma^{im}D_{i}A_m +  i \sum\limits_{m=4,5,6} \Gamma^{mn}[A_{m}, A_n] 
                  \right)   \varepsilon  \cr \cr
      =&&  \left( \Gamma^{ij}F_{ij} + 2 \Gamma^{i4}D_{i}A_4 + 2 \Gamma^{i5}D_{i}A_5
                     \right)      \varepsilon    \cr          \cr
         =&&  \left( \Gamma^{ij}F_{ij}  + 2 \Gamma^{i4}\sin \alpha (\half \epsilon_{ijk} F_{jk} )
          + 2 \Gamma^{i5} \cos \alpha  ( \half \epsilon_{ijk} F_{jk})   
                     \right)      \varepsilon              \cr     \cr
           =&&   F_{ij} \left( \Gamma^{ij} + \Gamma^{l4}\sin \alpha  \epsilon_{lij}  
          + \Gamma^{l5} \cos \alpha  \epsilon_{lij}   \right)      \varepsilon  \cr \cr
= && 2 \left[  F_{12} \left( \Gamma^{12} + \Gamma^{34}\sin \alpha 
          + \Gamma^{35} \cos \alpha   \right)   +   {\rm cyclic \; perm.} \{1 \rightarrow 2   \rightarrow 3\}          
          \right]   \varepsilon                                                    ~,
  \end{eqnarray} 
  leading to conserved supersymmetries defined through the equation:
  \begin{equation}
  \label{3dinstsusy1}
  [  \sin \alpha  \; \Gamma^{1234}   +    \cos \alpha  \;  \Gamma^{1235}  ]  \varepsilon =    + \varepsilon~,
  \end{equation}
  which is satisfied by half of the supersymmetries.    
  Few comments, parallel to the bosonic discussion of Sec.\ref{sec:bos},  are in order: 
\begin{enumerate}
\item For $\alpha={\pi\over 2}$,   we have:
  \begin{equation}
 \Gamma^{1234}  \varepsilon =  + \varepsilon  \qquad (\alpha={\pi\over 2}, \; {\rm self-dual \; BPS \;  monopole-instanton} )
\end{equation}
Thus,   $\varepsilon$ obeys the  same ``chirality" condition as the four dimensional instantons and the chirality of the 4d instanton zero modes and the BPS monopole-instantons are aligned.  These are the fermionic 
 zero modes associated with the monopole-instanton when the vev is aligned purely in  the $A_4$ direction.  
\item For  $\alpha=0$,  we have:  
  \begin{equation}
 \Gamma^{1235}  \varepsilon =  + \varepsilon  \qquad(\alpha=0,   \;  {\rm BPS \;  
 monopole-instanton} )
\end{equation} 
This is also the condition satisfied for a monopole particle in 4d, as well as its low-lying dyons (since the bosonic background for $\alpha = 0$ obeys the usual 4d monopole particle BPS equation, see discussion after (\ref{second})).  This condition will be particularly important when we consider Poisson-duality between the 4d-monopoles/dyons at large $\S^1$ and monopole instantons pertinent to small $\S^1$. 
 \item For $\alpha=-\pi/2$,  we have: 
  \begin{equation}
 \Gamma^{1234}  \varepsilon =  - \varepsilon  \qquad(\alpha=-\pi/2,  \; {\rm \overline{KK} \; anti-instanton} )
\end{equation}In this case,  $\varepsilon$ obeys the  same ``chirality" condition as the four dimensional anti-instantons. 
\end{enumerate}
If we now consider the more general BPS and $\overline {\rm KK}$ monopole  instantons whose action is given in  (\ref{small-L-action}), we need to generalize the equation obeyed by supersymmetries from (\ref{3dinstsusy1}) to:
  \begin{equation}
  [  \sin \alpha_{n_w}  \; \Gamma^{1234}   +    \cos \alpha_{n_w}   \;  \Gamma^{1235}  ]  \varepsilon_{n_w}  =    + \varepsilon_{n_w} 
  \label{ins-tower}
  \end{equation}
For general values of $a_4$ and $a_5$, the supersymmetries respected by the various instantons with different winding number  $n_w\in \Z$ are not the same.  However, in the regime $a_5 \gg a_4$, to leading order and for low-lying instantons in the tower, we can take: 
 \begin{equation}
   \sin \alpha_{n_w}   \approx 0, \qquad    \cos \alpha_{n_w}   \approx 1 ~.
  \end{equation}
Hence, for the low-lying band of the instanton tower, we have:
   \begin{equation}
\;  \Gamma^{1235}   \varepsilon_{n_w}  =    + \varepsilon_{n_w} , \qquad {\rm independent \; of } \; {n_w}  ~.
\label{ins-towerP}
  \end{equation}
  The fermion zero-mode wave functions can be found by applying a broken supersymmetry transformation (i.e. the one orthogonal to (\ref{ins-towerP})) to the monopole-instanton solution; we will not need the explicit form of these zero modes.

 \subsection{Monopole-dyon particle tower}

In the background of a monopole or dyon particle,  setting  $A_6$ to zero and using the first order  dyon differential equations (\ref{delta}) (thinking of $x^4$ as a compact Euclidean time direction, i.e., with $E_i = F_{4i}$), we obtain the preserved supersymmetries:
 \begin{eqnarray}
 \delta \Psi= &&( \Gamma^{MN}F_{MN}  )  \; \varepsilon \cr \cr
                   =&&  \left( \Gamma^{\mu\nu}F_{\mu \nu} + 2  \sum_{m=5,6} 
                  \Gamma^{\mu m}D_{\mu}A_m +  2 i \Gamma^{56}[A_{5}, A_6] 
                  \right)   \varepsilon  \qquad \cr \cr  
                   =&&  \left( \Gamma^{ij}F_{ij} + 2 \Gamma^{4i}F_{4i}          
      +  2 \Gamma^{i5}D_{i}A_5 
                     \right)      \varepsilon    \cr          \cr
                   =&&    \left( \Gamma^{ij}\epsilon_{ijk}B_k + 2 \Gamma^{4i}E_{i}          
      +  2 \Gamma^{i5}D_{i}A_5  
                     \right)      \varepsilon     \cr          \cr   
        =&&    D_i A_5  \left( \Gamma^{lm}\epsilon_{lmi} \cos \delta_{n_e} + 2 \Gamma^{4i} \sin \delta_{n_e}         
      +  2 \Gamma^{i5} \right)   \varepsilon 
                     \label{dyon-towerP1}                                                                    
  \end{eqnarray} 
For non-vanishing $D_i A_5$, the vanishing supersymmetry transformations can be found by multiplying the matrix equation in  parenthesis  with  $\Gamma^{5i}$ for a fixed  $i$. They, as usual, yield, the same equation independent of the value of $i$, given by:
  \begin{eqnarray}
&& \left( \cos \delta_{n_e}  \Gamma^{5ilm}\epsilon_{lmi} - 2 \sin \delta_{n_e}   \Gamma^{54}         
      +  2         \right)      \varepsilon_{n_e} =    + \varepsilon_{n_e}   \qquad  [i-{\rm fixed}]   \cr         \cr         
 && [  \cos \delta_{n_e}  \; \Gamma^{1235}   -   \sin \delta_{n_e}  \;  \Gamma^{45}  ]  \varepsilon_{n_e} =    + \varepsilon_{n_e}
       \label{dyon-towerP2}     
  \end{eqnarray}
For a generic dyon, the supersymmetries preserved are not aligned with the one of 3d-instantons (\ref{ins-tower}).  
However, in the limit when $\delta_{n_e} \sim 0$: 
    \begin{equation}
\;  \Gamma^{1235}   \varepsilon_{n_e} =    + \varepsilon_{n_e}, \qquad {\rm independent \; of } \; {n_e} ~,
\label{dyon-towerP}
  \end{equation}
 as in (\ref{ins-towerP}).  The $\delta_{n_e} \sim 0$ condition, and hence (\ref{dyon-towerP}) holds for the band of states for which $n_e^2 \ll {4 \pi^2 \over g_4^2}$, see eqn.~(\ref{delta}). This guarantees the working of the approximate version Poisson duality  (\ref{Poissond})  even when fermions are incorporated. 
 Similar to the remarks after (\ref{ins-towerP}), the fermion zero modes can be found by applying a broken supersymmetry transformation to the bosonic solution.

 \section{$\mathbf{SU(N)}$ generalization of Poisson duality}
 \label{sungeneral}
We now turn to $SU(N)$, $N\geq 3$, generalization of the Poisson duality between the 4d monopole/dyon tower and 3d instanton tower. 

On $\R^4$, the set of   vacua of the classical gauge theory is the space of commuting covariantly  constant scalars $[ \phi, \phi^{\dagger}] = 0$, or $[A_5 , A_6]=0.$ For convenience,  we  take  $A_6=0$ and $A_5=   {\rm diag} [v_1, v_2, \ldots, v_N]$, and we choose a point in the moduli space where the long distance theory fully abelianizes, 
$SU(N) \rightarrow U(1)^{N-1}$.
There  are $N-1$ types of monopole particles  whose charges are proportional to the $N-1$ simple roots $ \alpha_i$   of the Lie algebra, of magnetic charges $ Q_i = 4 \pi  \alpha_i$, $i=1, \ldots N-1$.
The mass spectrum of  these $N-1$ type of lightest  monopole particles and their dyonic tower is: 
\begin{eqnarray}
M_{i}  \equiv M_{i, i+1}=|v_i -v_{i+1}|  \sqrt{ \left( \frac{4 \pi}{g_4^2} \right)^2 n_m^2 + n_e^2} 
  \label{bpsmass3}
 \end{eqnarray}
For $n_m=0, n_e=1$,  eqn.~(\ref{bpsmass3}) reduces to the mass formula for the  lightest W-bosons   due to adjoint 
Higgsing,  $M_{i}=|v_i -v_{i+1}|$. For $n_e=0, n_m=1$,  (\ref{bpsmass3}) is  the semi-classical mass formula for the monopoles, $M_{i} = |v_i -v_{i+1}|   \left( \frac{4 \pi}{g_4^2} \right)$.  Our primary interest are the set of monopoles with $n_m=1$ and arbitrary electric charge 
$\sim n_e \alpha_i$.

For the theory   compactified on  $\R^3 \times \S^1$,  
the Wilson line along the compact direction $\Omega= e^{i \int_{S^1} A_4}$  can  also be  interpreted as scalar from the lower dimensional point of view. However, $A_4$ differs from $A_5$ and $A_6$ since it is an angular variable, and as in our $SU(2)$ example, this plays a crucial role. 
The vacua of  the classical gauge theory on $\R^3 \times \S^1$ are spanned by the space of commuting covariantly constant scalars: 
\begin{equation}
    [ A_{3+i}, A_{3+j}]=0, \;\;\;  i, j=1, 2,3 \,.
\end{equation}
which may be  parameterized as  three sets of $N$ eigenvalues,
\begin{subequations}
\begin{align}
    A_4 &= 
    \frac{1}{L} {\rm diag}
    \left( \omega_1 ,\, \omega_2,\, \cdots,\, \omega_N \right) , \qquad \sum_{i=1}^{N} \omega_i=0 \;\; {\rm mod} \; 2 \pi ~,
\\
   A_5 &=
    {\rm diag}
    \left( v_1, \, v_2 ,\, \cdots,\, v_N \right) , \qquad \;\;  \sum_{i=1}^{N} v_i=0 ~,
    \end{align}
\label{backg}
\end{subequations}

For convenience, we  set  $A_6=0$. For generic configurations with distinct eigenvalues,  the long distance theory fully abelianizes. At such  points, 
the subgroup of global gauge transformations which preserve 
the diagonalized form (\ref{backg})
is the Weyl group $\mathcal W$ of $SU(N)$, whose elements
\emph{simultaneously} permute the eigenvalues of the two scalars,
$(L^{-1}\omega_i, v_i) \rightarrow (L^{-1} \omega_{\sigma(i)}, v_{\sigma(i)}) $,
where $\sigma \in \SN$ ($\SN$ is the $N$-element permutation group).   In this sense, 
$\SN$  acts on $N$-``eigenbranes"  whose positions are given by:
\begin{equation}
{\bf r}_i= (L^{-1}\omega_i, v_i)  
\end{equation} 
as ${\bf r}_i \rightarrow {\bf r}_{\sigma (i)}$. Due to periodicity of   $\omega_i$, the images of these  eigenbranes are located at: 
\begin{equation}
{\bf r}_i (n_w)= (L^{-1}(\omega_i + 2 \pi n_w), v_i) \; , \qquad  n_w \in \Z~.
\end{equation} 
The utility of this description is that it geometrizes various aspects of monopole-instantons at  small $\S^1$. 
The action of the monopole-instantons associated with magnetic charge $\alpha_i$  embedded into $\R^3 \times \S^1$ is: 
 \begin{eqnarray}
 S_i=&&  \frac{ 4 \pi L }{g_4^2}     | {\bf r}_i (0) -{\bf r}_{i+1} (0) |  = 
  \frac{ 4 \pi L }{g_4^2}  \sqrt{(v_i -v_{i+1})^2 + L^{-2} \left( \omega_i - \omega_{i+1} \right)^2}  
  \label{small-L-I}
 \end{eqnarray}
and the action  of the winding 3d instantons (the  Kaluza-Klein tower of  (\ref{small-L-I})) is:
 \begin{eqnarray}
 S_{i, n_w}=&&  \frac{L }{g_4^2}     | {\bf r}_i (0) -{\bf r}_{i+1} (n_w) | 
 \cr  = &&
  \frac{4 \pi L }{g_4^2}  \sqrt{(v_i -v_{i+1})^2 +  L^{-2} \left( \omega_i - \omega_{i+1} + 
 2 \pi n_w  \right)^2}    ,   \qquad 
  n_w \in \Z~.
  \label{small-L-II}
 \end{eqnarray}
This formula is noting but the distance between ${\bf r}_i (0)$ and the image of 
${\bf r}_{i+1}(0)$ labeled as      ${\bf r}_{i+1}(n_w)$

The  fugacity function   associated with monopole-instantons of charge $\alpha_i$ is therefore, similar to  eqn.~(\ref{fugacity}): 
   \begin{eqnarray}
  F_i (\omega_i - \omega_{i+1})=  \sum_{{n_w}  \in \Z} e^{ -  S_{i, n_w}}  =  \sum_{{n_w}  \in \Z} e^{  -   \frac{4 \pi L }{g_4^2}  \sqrt{(v_i -v_{i+1})^2 +  L^{-2} \left( \omega_i - \omega_{i+1} + 
 2 \pi n_w  \right)^2}  }~.
 \label{function1}
  \end{eqnarray} 
  We should  note that  all the instantons  in the   tower have the same magnetic charge, but their topological charges differ by one unit. The part of the sum contributing to  
$n_w\geq 0$  can  smoothly be deformed to  the  self-dual monopole-instantons by taking 
  $(v_i -v_{i+1})=0$, whereas under the same smooth deformation,  the   $n_w\leq -1$ terms arise from      non-selfdual monopole-instantons.  As it was the case in $SU(2)$, the sum over the winding number  combines  self-dual and non-selfdual monopole-instantons into a single tower. In the sense of magnetic charge, this is possible because the simple  roots $ \{ \alpha_1,  \alpha_{2}, \ldots,   \alpha_{N-1}\} $ and affine  root $ \{ \alpha_N\}$  of   $SU(N)$ satisfy  
$  \alpha_i = \alpha_i + n_w  \sum_{j=1}^{N} \alpha_j $. 
     
Since the function $ F_i (\omega_i - \omega_{i+1})$ from (\ref{function1}) is periodic in $\omega_i - \omega_{i+1}$, we can Fourier expand it and follow the  steps given in Section \ref{poissongeneral}. The result is 
  \begin{eqnarray}
 F_i (\omega_i - \omega_{i+1}) =    \sum_{n_e \in \Z}    \frac{vL}{ \pi}  \cos\delta_{n_e} \; \;   K_1 (LM(1, n_e)) e^{ i n_e (\omega_i - \omega_{i+1}) }  \; .
\label{sqrt21}
   \end{eqnarray} 
where $M(1, n_e)$ is the mass formula for dyon particles given in (\ref{bpsmass3}).  
This is the generalization of Poisson duality between the   tower of  3d monopole-instantons and the 4d tower of dyonic excitations of a monopole.

\end{document}